\documentstyle[12pt,aasms4]{article}
\journalid{}{}
\articleid{}{}

\begin{document}
\title{\bf \LARGE The HI and Ionized Gas Disk of the Seyfert Galaxy
NGC 1144 = Arp 118: A Violently Interacting Galaxy with Peculiar
Kinematics}

\author{M. A. Bransford\altaffilmark{1,2} \& P. N. Appleton\altaffilmark{1}}
\authoremail{mabransf@pandora.nmsu.edu, pnapplet@iastate.edu}
\affil{Erwin W. Fick Observatory and Department of Physics and
Astronomy, Iowa State University, Ames, IA 50011}

\author{C. F. McCain \& K. C. Freeman}
\authoremail{kcf@mso.anu.edu}
\affil{Mount Stromlo and Siding Springs Observatories, Australian National
     University, Private Bag, Weston Creek P.O., ACT 2611, Australia}
\altaffiltext{1}{Visiting astronomer at the National Radio Astronomy 
Observatory. NRAO is a facility of the National Science Foundation
operated under cooperative agreement by Associated Universities, Inc.}

\altaffiltext{2}{Current address: Department of Astronomy, New Mexico
State University, Las Cruces, NM 88003-8001}

\begin{abstract}

We present observations of the distribution and kinematics of neutral
and ionized gas in NGC 1144, a galaxy that forms part of the Arp 118
system. Ionized gas is present over a huge spread in velocity (1100 km
s$^{\rm -1}$) in the disk of NGC 1144, but HI emission is detected
over only $\onethird$ of this velocity range, in an area that
corresponds to the NW half of the disk. In the nuclear region of NGC
1144, a jump in velocity in the ionized gas component of 600 km
s$^{\rm -1}$ is observed. Faint, narrow HI absorption lines are also
detected against radio sources in the SE part of the disk of NGC 1144,
which includes regions of massive star formation and a Seyfert
nucleus. The peculiar HI distribution, which is concentrated in the NW
disk, seems to be the inverse of the molecular distribution which is
concentrated in the SE disk. Although this may partly be the result of
the destruction of HI clouds in the SE disk, there is circumstantial
evidence that the entire HI emission spectrum of NGC 1144 is affected
by a deep nuclear absorption line covering a range of 600 km s$^{\rm
-1}$, and is likely blueshifted with respect to the nucleus. In this
picture, a high column-density HI stream is associated with the
nuclear ionized gas velocity discontinuity, and the absorption
effectively masks any HI emission that would be present in the SE disk
of NGC 1144.

\end{abstract}


\section{Introduction}

The Arp 118 system is comprised of a distorted disk galaxy NGC 1144
and an elliptical galaxy NGC 1143. NGC 1143 is $\sim$40$\arcsec$ to
the NW of NGC 1144, which corresponds to a linear separation of $\sim$20 
kpc (assuming a distance of 110 Mpc to Arp 118).  NGC 1144 is also
classified as a Seyfert 2 galaxy (Hippelein 1989, hereafter
H89). There exists a knotty ``ring'' or loop, that extends from NGC
1144 towards the companion in the north-west, and this is most easily
seen in the H$\alpha$ (+continuum) image of the galaxy pair shown in
Figure 1. However, the structure of the disk of NGC 1144 is more
complicated than a simple ring, containing many loops and filaments
and a prominent dust-lane which runs from the SE to the NW from a
point east of the nucleus. These details can be seen more clearly in
the red HST image of Figure 1 (image courtesy of M. Malkan).

The inner region of the disk of NGC 1144 contains extended regions of
star formation. Quite apart from the strong H$\alpha$ emission from
these regions (H89), Joy \& Ghigo (1988) showed that a giant H~II
region complex NW of the nucleus contributed 35$\%$ of the galaxy's
substantial $\lambda$10$\mu$m emission. They estimated that NGC 1144
has a bolometric luminosity of 2.5 $\times$ 10$^{\rm 11}$ L$_{\odot}$,
80$\%$ of which is re-radiated in the thermal infrared. Their 6 and 20
cm radio continuum images revealed several radio sources corresponding
with blue optical emission knots, suggesting that the radio continuum
was originating from recent star formation. Higher sensitivity VLA
radio observations by Condon et al. (1990) supported the idea that, in
addition to the Seyfert nucleus, there are other regions of bright
radio emission within the inner disk, including a strong source to the
east, and a spiral-like arc of emission extending over several
kiloparsecs.

The early pioneering mapping of the H$\alpha$ velocity field of NGC
1144 showed a huge velocity spread across the galaxy of over
$\sim$1100 km s$^{\rm -1}$, implying the necessity for a mass for the
galaxy in excess of 10$^{\rm 12}$ M$_{\odot}$ (H89). Furthermore, the
velocity-field is highly asymmetric, with a change in velocity of over
700 km s$^{\rm -1}$ from the center of NGC 1144 to the south-eastern
edge, but only a change of over 400 km s$^{\rm -1}$ from the center to
the north-western edge of the disk.  The very large spread in velocity
seen in the ionized gas is also apparent in the molecular CO emission
(Gao, Solomon, Downes, \& Radford 1997, hereafter GSDR), where CO
linewidths of 1100 km s$^{\rm -1}$ are seen in their IRAM single-dish
(22$\arcsec$ beam) spectra. These authors showed Arp 118 to have a CO
luminosity nearly twice that of Arp 220. The CO emission is
distributed non-uniformly, and is highly concentrated in the SE
quadrant of the ring, with much weaker emission from the NW quadrant.
CO maps made with the IRAM interferometer (5$\farcs$3 $\times$
2$\farcs$5 beam) revealed that the CO emission is not centrally
concentrated in the nucleus, but traces the southern arm of an inner
ring coincident with the most luminous H~II regions.

Hippelein (1989) suggested a dynamical model for a ring-making
collision between NGC 1144 and NGC 1143 in which the disk of NGC 1144
is severely distorted by the collision. However, in order to reproduce
the large spread in velocity, the mass of NGC 1144 seemed excessive
(of the order of 10$^{\rm 12}$ M$_{\odot}$). More recently, Lamb,
Heran \& Gao (1998) have attempted to model the interaction in a more
sophisticated manner, including a massive halo, full self-gravity and
the inclusion of gas-dynamics. Although their final model, with a
total mass (including massive halo) for NGC 1144 of 5 $\times$
10$^{\rm 11}$ M$_{\odot}$ does produce a large velocity spread (of
about 950 km s$^{\rm -1}$) in the disk after the collision, it fails
to reproduce the asymmetry in the velocity field of the disk,
discussed earlier.

This paper presents VLA HI observations of Arp 118 in an attempt to
shed new light on the interpretation of the dynamics, as well as
presenting new H$\alpha$ observations, revealing the kinematics of the
ionized gas in the disk of NGC 1144, using the ANU 2.3-m telescope.
We assume throughout this paper a distance to Arp 118 of 110 Mpc,
based on an assumed heliocentric velocity\footnote{This value of the
assumed radial velocity is based on an interpretation of H$\alpha$
isovelocity map presented in this paper. As we shall see, emission
lines across the nucleus reveal a sudden jump in velocity which is
probably related to a peculiar outflow--see text.} for the galaxy of
8800 km s$^{\rm -1}$, and a Hubble constant of 80 km s$^{\rm -1}$.

\section{The Observations}

\subsection{The HI Observations of Arp 118}

Arp 118 was observed at the VLA on 17 March 1996 using all 27
telescopes configured in the C-array. The correlator was set in a two
IF (intermediate frequency) mode (2AD) with on-line Hanning smoothing
and 32 channels per IF. For each IF we used a bandwidth of 6.25 MHz,
which provided a frequency separation of 195.3 kHz per channel. This
frequency separation corresponds to a velocity separation of 43.8 km
s$^{\rm -1}$ channel$^{\rm -1}$ in the rest frame of the galaxy (using
the optical definition of redshift). Arp 118 contains H~II regions (in
the ring) that have radial velocities spanning a range of more than
1000 km s$^{\rm -1}$ (H89). GSDR reported a wide velocity range in CO
emission (1100 km s$^{\rm -1}$). In order to achieve velocity coverage
consistent with these observed velocity ranges, IF1 was centered at
8269 km s$^{\rm -1}$ and IF2 at 9230 km s$^{\rm -1}$. The resulting
velocity coverage of our observations was 2275 km s$^{\rm -1}$. A
total of 3 hours and 47 minutes were spent on source. Strong winds
during the observations were responsible for a loss of 25$\%$ of our
original allocation of observing time. Flux and phase calibration were
performed using the sources 3C48 and 0320+053 (B1950), respectively.

The data were first amplitude and phase calibrated, and bad data due to
interference were flagged and ignored by the AIPS software. An image
cube was created from the UV data by giving giving more weight to those
baselines that sampled the UV plane more frequently (natural weighting).
This provided a synthesized beam with a FWHM of 21$\farcs$2 $\times$
17$\farcs$0.

Subtraction of the continuum emission in each line map was performed
using a standard interpolation procedure based on five continuum maps
free from line emission at the ends of the bands. The rms noise per
channel was 0.36 mJy beam$^{\rm -1}$. The highest dynamic range achieved
in any channel map was 10:1. A single image cube (we will refer to this
as the combined cube) was made by combining IF1 and IF2.

We determined the HI line integral profiles, or moments, from the
spectral data cube. The zeroth moment corresponds to the integrated
intensity over velocity and the first moment to the intensity weighted
velocity (the second moment--the velocity dispersion map--was
uninteresting and was not included here). We employed the following
procedure. We smoothed the combined cube with a beam twice that of the
synthesized beam. As a result, a smooth cube was created with a
resolution of 42$\farcs$4 $\times$ 34$\farcs$0. We applied a
signal-to-noise threshold to the smoothed cube, blanking all pixels
that fell below a 3$\sigma$ cutoff. A new image cube was then created
by applying the blanked, smoothed cube as a ``mask'' to the original
full resolution map. The total HI surface density map was then
produced from the full resolution, blanked cube. The first-moment
(intensity-weighted mean velocity field) was created from the same
blanked cube.  For those channels where obvious absorption was
present, we employed a different procedure. In those cases, we
directly fitted to the channel maps showing clear absorption using the
AIPS routine IMFIT. The depth of the absorption was then determined
from the fit.

\subsection{Optical Spectroscopy}

Observations of Arp 118 were made in 1992 with the Double Beam
Spectrograph on the 2.3-m ANU telescope at Siding Spring Observatory.
Longslit spectra of length 400$\arcsec$ were obtained at a series of
position angles across the disk of NGC 1144. The dispersion was 25 km
s$^{-1}$ pixel$^{-1}$ and the slit width was 1$\farcs$8, which
projects to 2 pixels at the detector. The positioning of the slits
will be discussed in a later section. A full discussion of the
spectral reduction and a more complete discussion of these optical
data will be discussed in a separate paper. In the present paper, we
will restrict our discussion to the presentation of the average
velocity field of the H$\alpha$ emission and its ramifications for the
interpretation of the HI observations.

\section{Integrated Properties of the HI, its Distribution and the
Relationship Between the HI and CO Emission}

Figure 2 shows the global HI profile of Arp 118, obtained by
integrating the HI emission spatially over the galaxy at each
velocity.  Notice the narrow spread in velocity of the emission,
spanning $\sim$350 km s$^{\rm -1}$, compared with the huge velocity
spread seen in the ionized and molecular gas. With the exception of a
faint feature seen at 9200 km s$^{\rm -1}$, the global HI profile has
about one-third the velocity spread of the CO and H$\alpha$ emission,
as shown in Figure 3.  A majority of the CO emission is at high
velocities (in the SE) whereas the HI emission is observed at low
velocities (in the NW). Taken together, the total spread in velocity
seen in the HI and CO (from 8180-9290 km s$^{\rm -1}$) agrees closely
with the spread seen in the ionized gas (H89, also this paper).

Two faint but narrow HI absorption features may be present at
velocities centered around 8800 and 9000 km s$^{\rm -1}$. The
lower-velocity line is close to the noise-level of the observations,
but the second is stronger and is seen at the 3-5$\sigma$ level over
three channels (see also discussion in $\S$4).  Both lines, if real,
appear spatially as negative contours against the brightest part of
the radio continuum source seen in the galaxy (see $\S$4 and
6.2). There is also fainter emission seen in the higher velocity
channels near 9200-9300 km s$^{\rm -1}$ (see $\S$4). We will discuss
these absorption lines and the possibility that much of the HI profile
from NGC 1144 is nullified by a stronger, and broader, absorption line
in $\S$6.

Integrated HI properties for NGC 1144 are given in Table 1. Table 1
also includes the integrated HI properties for a dwarf galaxy KUG
0253-003, located $\sim$8$\arcmin$ NE of Arp 118 (which we discuss
fully in $\S$7). The total HI mass was determined from the spectrum
of Figure 2 from the formula:

\begin{equation}
{\rm M_{H} = 2.356 \times 10^{5}~D^{2} F_{H}~~~(M_{\odot}),}
\end{equation}

\hspace{-2em}where D is the distance in Mpc, and 

\begin{equation}
{\rm F_{H} = \int_{}^{} S_{\nu}~dV~~~~~(Jy~km~s^{-1}).}
\end{equation}

Two independent single-dish profiles for NGC 1144 are available from
the literature. The integrated flux density F$_{H}$ detected by
Bushouse (1987) using the NRAO 91-m telescope was 2.83 Jy km s$^{\rm
-1}$ and by Jeske (1986) using Arecibo was 2.55 Jy km s$^{\rm
-1}$. Our value was determined to be 2.40 Jy km s$^{\rm -1}$, or 85$\%$
of the total emission quoted by Bushouse and 94$\%$ of the total
emission quoted by Jeske. The lower value obtained here is consistent
with the possibility that the C-array has resolved-out faint extended
HI emission which would be detected by the single-dish observations.
However, neither single dish spectra cover the velocity range of the
possible absorption lines, and our estimate of the total HI mass given
in Table 1 does not take into account the possible contamination of
the HI emission line profile by a broad, but deep, HI absorption
feature.

The total HI mass of Arp 118 based on our C-array observations is
M$_{\rm H}$ = 7.0 $\times$ 10$^{\rm 9}$ M$_{\odot}$. This compares
with an estimated molecular mass of 1.8 $\times$ 10$^{\rm 10}$
M$_{\odot}$ (GSDR), based on the standard galactic H$_{\rm 2}$
mass-to-CO luminosity. GSDR, however, noted that the standard
conversion factor might be 2-3 times lower in Arp 118 than in the
Milky Way, which would therefore correspond to a molecular hydrogen
mass of $\sim$6-9 $\times$ 10$^{\rm 9}$ M$_{\rm \odot}$.

We also include in Table 1 an estimate of the dynamical mass M$_{\rm
d}$ (see for example Appleton, Charmandaris, \& Struck 1996) of NGC
1144. The value of M$_{\rm d}$ was calculated using the formula:

\begin{equation}
{\rm M_{d}~~=~~\frac{[(\onehalf)\Delta V_{\scriptsize \onehalf}
cosec(i)]^{2}~R_{HI}}{G}~~~~~~(M_{\odot})}
\end{equation}

\hspace{-2em}where R$_{\rm HI}$ is the radius of the HI disk,
$\Delta$V$_{\scriptsize \onehalf}$ is the width of the profile where
the flux density is one-half the peak, and i is the inclination of the
disk. This formula applies to gas that is in bound circular orbits.
Note that the assumption of circular orbits and a ``normal'' rotation
curve may be quite incorrect in Arp 118.  Because the spread in
velocities of the HI emission is about one-third that of the CO (and
H$\alpha$), we have estimated $\Delta$V$_{\scriptsize \onehalf}$ from
the combined HI and CO profiles, giving a $\Delta$V$^{\rm
CO+HI}_{\scriptsize \onehalf}$ of 1020 km s$^{\rm -1}$. Based on the
HI map, we estimate R$_{\rm HI}$ to be 10 kpc, and i = 50$\arcdeg$,
based on the ratio of the major to minor axis of the optical outer
ring. Hence we obtained a value of M$_d$~=~1 $\times$ 10$^{\rm 12}$
M$_{\odot}$.

Figure 4 contains the integrated HI distribution of Arp 118, overlayed
with a digitized sky survey (DSS) grey-scale image of Arp 118. The HI
emission is distributed non-uniformly throughout the disk of NGC 1144,
and is concentrated in the NW quadrant of the ring, with fainter
emission to the south.

Figure 5 (see also Figure 6) is an overlay of the integrated HI map
with the a grey-scale image of the CO knots imaged by GSDR. As
mentioned, the high resolution CO data of GSDR shows that the CO
emission traces the southern arm of the radial ring wave (the
approximate location of the ring has been drawn to guide the
eye). Further, the knots of CO emission are coincident with luminous
H~II regions, also concentrated along the southern arm of the
ring. Inspection of Figure 5 reveals that the brightest complexes of
CO emission do not overlap the HI emission.  In Figure 6 we have
convolved the CO emission to the same beam size as our VLA data. The
CO emission appears to, more strikingly, ``fill-in'' the ``missing''
HI emission in the south-eastern half of NGC 1144. Considering Figures
3, 5, and 6, the HI and CO appear ``segregated'', both spatially and
kinematically. Note, however, that we use the word ``segregated''
rather loosely, since there is weaker CO emission in the NW quadrant
(see next paragraph), and low-level HI emission in the SE. The word is
used to point to the fact that the distributions of CO and HI are
asymmetrically peaked, CO to the SE and HI to the NW.

Interestingly, there is a striking similarity between the single dish CO
profile in the NW quadrant of the of the ring and our HI profile (see
Figure 1 in GSDR for the single dish profile). The CO profile peaks at a
velocity of 8250 km s$^{\rm -1}$ and the HI profile peaks at 8267 km
s$^{\rm -1}$. Both profiles fall off rapidly towards higher velocities.

\section{HI Kinematics}

In Figure 7 we present those channel maps containing detectable
emission from 8180 to 8661 km s$^{\rm -1}$. At the rest frame of the
galaxy, the channel separation is 43.8 km s$^{\rm -1}$.  The emission
at the lower velocity channels is mainly concentrated in the northern
part of the galaxy in the region associated with the NW quadrant of
the outer optical ring seen in Figure 1, but also extends
significantly to the NE, outside the optical bounds of the galaxy (V =
8267 km s$^{\rm -1}$). The emission becomes more scattered and moves
further south as one proceeds to higher velocities. There is a
marginal detection of faint emission to the SE of the main body of the
galaxy around 8573 km s$^{\rm -1}$ and 8661 km s$^{\rm -1}$.  With the
exception of a faint HI features centered on the nucleus at velocities
of 9200 to 9300 km s$^{\rm -1}$ (see later), no significant HI
emission is observed at velocities in excess of 8661 km s$^{\rm -1}$.

Figure 8 contains the mean velocity field of Arp 118, which appears
rather disturbed as compared with the optical velocity (see below).
The velocities appear, approximately, to run from lower velocities in
the north to higher velocities in the south.  Except for contours at
8355 km s$^{\rm -1}$ , there is little emission from the SE disk of
NGC 1144. The velocity isocontours generally have a bent and/or
rather distorted S-shaped appearance. In the regions of common
coverage between the optical and HI velocity fields (see later), there
is approximate agreement. However, the fact that much of the HI disk
of NGC 1144 is missing in the SE make the HI velocity field very
difficult to interpret.

Figure 9 contains another set of channel maps showing the faint
absorption features seen against the nuclear radio disk.  The
absorption appears in 5 channels, at the same spatial location, but is
strongest at a velocity of 9011 km s$^{\rm -1}$. Figure 9 also
contains a faint emission feature seen near the nuclear region. This
emission appears in 5 channels, most strongly in in the 9318 and 9274
km s$^{\rm -1}$ channels.

In order to show the absorption features more clearly, we have
constructed, in Figure 10, a spectrum through the unblanked data cube
centered on the galaxy's nucleus. This has the effect of integrating
{\it only} that emission or absorption associated with the nuclear
source. This spectrum differs in nature from the integrated (global)
spectrum of Figure 2 because it concentrates on the nuclear regions
only, and is determined from the cube without any flux-threshold
blanking. Figure 10 reveals faint nuclear absorption and emission.
Interestingly, Figure 10 suggests that the two faint absorption
features, which we isolated in Figure 2, may be part of a faint,
broader absorption complex. The absorption features at their deepest
are detected at the 4$\sigma$ (8791 km s$^{\rm -1}$) and 5$\sigma$
(9011 km s$^{\rm -1}$) level. Figure 10 also contains the faint
emission features at velocities of 9200 to 9300 km s$^{\rm -1}$, as
shown in Figure 2. These emission features are at a similar level of
significance as the absorption features. In $\S$6.2 we will return to
the astrophysical implications of the broad absorption features and
their possible effect on the total HI profile.

\section{The Kinematics of the Ionized Disk}

In Figure 11, we show the mean velocity field of the H$\alpha$
emitting gas based on the observations made with the ANU 2.3-m
telescope. A more complete discussion of these optical data will be
presented in a separate article (McCain and Freeman, in
preparation). In this paper we restrict ourselves to a discussion of
the velocity field of the galaxy. In the upper right of Figure 11 we
show the coverage of the long-slits used to create the velocity
map. The slits ranged in position angle from 318$\arcdeg$ to
273$\arcdeg$, and each slit was positioned on the nucleus of the
elliptical companion NGC 1143 during each observation. The one
exception was a slit oriented at 309$\arcdeg$ which was positioned to
pass through the nucleus of NGC 1144, but was not centered on NGC
1143. Ionized gas emission was detected over the entire face of NGC
1144. The velocity field in the upper left of Figure 11 shows the
entire H$\alpha$ emitting disk, and the inset to the lower right shows
the details of the disk closer to the center. We overlay the contour
map of the velocity field with a grey-scale representation of the
$\lambda$20cm radio continuum emission from Condon et al. (1990), and
we will discuss the significance of this later.

The velocity field of NGC 1144, derived from the ANU 2.3-m spectra, is
very similar in overall appearance to the earlier work of H89. The
gross characteristics of the system are large-scale rotation over a
velocity range of 1100 km s$^{\rm -1}$. The ANU 2.3-m data show a
sudden velocity discontinuity in the vicinity of the nucleus, which
in the inset shows a drop in radial velocity of about 600 km s$^{\rm
-1}$, from a value in the disk of around 9000 km s$^{\rm -1}$ to a
value around 8400 km s$^{\rm -1}$. This region is almost unresolved at
the level of the seeing (1$\arcsec$), and so the velocity
discontinuity is most probably associated with the nucleus. The cross
shows the nominal position of the optical nucleus, but this is not
well determined from the spectra. (Note: we have assumed in overlaying
the map with the radio image that the velocity discontinuity
corresponds to the position of the nuclear source).

Putting aside for the moment the question of the origin of the
velocity jump at the center of the galaxy, we now address the question
of the value of the systemic velocity of the galaxy. Based on the
appearance of the isovelocity contours, it would be normal to consider
the systemic velocity of a galaxy to be the velocity associated with
the contour which passed through the nucleus. Ignoring the velocity
discontinuity, the most natural contour which would pass close to the
nucleus, and which runs parallel to the minor axis of the galaxy,
would be the velocity contour at 9050$\pm$50 km s$^{\rm -1}$. It is
interesting to note that the ANU 2.3-m spectra show CaII H and K
absorption lines, presumably from the underlying stellar population in
the galaxy, which yield a velocity of 9000$\pm$100km s$^{\rm -1}$ for
the nucleus, in approximate agreement with the global velocity field
of the ionized gas. It is clear that the velocity discontinuity is a
major deviation away from this value for the systemic velocity, since
the emission lines in the nucleus yield a velocity of 8400km s$^{\rm
-1}$. Does this mean that the nucleus is moving relative to the gas
disk, or simply that there are peculiar motions in the ionized gas in
and around the core? In the next section, we argue for the latter,
based on the HI absorption results.

\section{What is the Origin of the Asymmetric HI Distribution in NGC 1144?}

One of the surprising results of our HI observations of NGC 1144 is
the rather dramatic asymmetry in the distribution of the gas in the
disk. Our VLA observations show that most of the gas detected is
centered on the NW half of the galaxy, whereas ionized and molecular
gas are spread over the whole disk, but are concentrated mainly in the
southeastern half of the galaxy. This apparent segregation is also
reflected in the kinematics. HI from the disk of NGC 1144 is seen
exclusively at velocities below 8600 km s$^{\rm -1}$, and little
detectable HI emission is seen over a wide range of velocities
represented by a large part of the ionized and molecular disk. We
address below two possible explanations for the missing HI. The first
is the possibility that the segregation is a result of the large-scale
conversion of HI into molecules via strong shocks in the disk of this
violently colliding galaxy. The second, and perhaps more plausible
explanation, is that HI is present over the entire velocity range, but
that very powerful, and extremely broad HI absorption lines are
present in the nuclear regions that have ``nullified'' the entire
emission-line profile of the galaxy over a 600 km s$^{\rm -1}$
interval. The evidence for the latter is strengthened by the discovery
of a rapid change in velocity in the ionized gas over the same
velocity range in the nucleus, as we discussed in the previous
section.

\subsection{Conversion of atomic to molecular gas?}

If we take at face-value the asymmetry in the HI distribution in the
disk of NGC 1144, we can attempt to determine quantitatively the
relative importance of the atomic and molecular hydrogen content
across the disk NGC 1144 by calculating the molecular-to-atomic gas
mass ratio, M$_{\rm H_{\rm 2}}$/M$_{\rm HI}$, in the NW and SE
regions. We have converted the observed CO emission from GSDR into a
mass of molecular gas (using the standard galactic conversion from CO
to H$_{\rm 2}$), and calculated the HI mass, separately, for the NW
and SE regions. The molecular-to-atomic gas mass ratio is 0.77 in the
NW section of the ring, consistent with the observed ratios in normal
spiral galaxies (Casoli et al. 1998). The ratio is very large in the
SE section, 17.6, consistent with ratios measured in other
interacting, infrared luminous galaxies (Mirabel \& Sanders 1989). If
this scenario is correct, the gas mass ratios suggest that the
dominant state of the interstellar medium proceeds from mainly atomic
in the NW, to molecular in the SE. In the SE region less than 6$\%$ of
the interstellar gas is in atomic form. Considering that the total
(HI+H$_{\rm 2}$) mass of hydrogen remains approximately equal in both
regions ($\sim$1 and 2 $\times$ 10$^{\rm 10}$ M$_{\odot}$ in the NW
and SE, respectively) suggests the possibility that there may be a
large scale conversion of HI to H$_{\rm 2}$ in the southern region of
NGC 1144.

One possible way to enhance the molecular gas mass fraction is by
compressing the disk in NGC 1144. Modest compression and/or shocks in
the interstellar medium can lead to the conversion of atomic to
molecular gas (Elmegreen 1993; Honma, Sofue, Arimoto 1995). The
salient features of Elmegreen's model are that the HI-H$_{\rm 2}$ gas
phase transition depends sensitively on the pressure in the
interstellar medium and the radiation field: H$_{\rm 2}$ molecules are
formed on the surface of dust, but can be destroyed by UV photons. The
results of his models imply that large regions within galaxies can
spontaneously convert atomic into molecular gas following an
interaction, a tidal encounter that has led to accretion, or increase
in the gas surface density.

While it is plausible that such a process is occurring in the disk of
NGC 1144, there exists a much more likely explanation for the
``missing'' HI, namely the possibility of HI absorption.

\subsection{HI absorption}

We will begin this section by discussing the HI absorption line seen
at approximately 9000 km s$^{\rm -1}$ in the channel maps, as
shown in Figure 9. Figure 12 contains a contour map of the continuum
emission, overlayed upon a grey scale representation of the integrated
HI emission. The absorption appears nearly coincident with the peak
continuum emission in NGC 1144. The total flux we detected from the
continuum source associated with NGC 1144 was 139.5 mJy, within a
synthesized beam of 21$\farcs$2 $\times$ 17$\farcs$0, which is in
excellent agreement with the 20 cm flux of 136 mJy from Condon et
al. (1990) in an 18$\arcsec$ $\times$ 18$\arcsec$ beam. However,
higher resolution observations of the continuum emission by Condon et
al. reveal a compact nucleus, and two extra-nuclear emission regions,
as well as a faint arc within the inner disk (see greyscale overlay in
upper right of Figure 11). In determining the HI column density
responsible for the faint absorption line at 9000 km s$^{\rm -1}$ it
is necessary to assume a plausible location for the HI absorbers,
i.e., which of the continuum emitting regions, seen in the higher
resolution continuum image, the absorbers cover.

For an HI cloud seen in absorption (F$_{\rm abs}$) against a continuum
source (F$_{\rm con}$), the optical depth, $\tau$, of the cloud is
given by (e.g. Mirabel 1982):

\begin{equation}
{\rm \tau~~=~~ln(1~+~\frac{F_{abs}}{F_{con}})}
\end{equation}

\hspace{-2em}and hence the hydrogen column density N$_{\rm HI}$ is given by:

\begin{equation}
{\rm N_{HI} = 1.823 \times 10^{20}~\int^{}_{}~T_{S100} ~\tau~dv~~~atoms~cm^{-2}}
\end{equation}

\hspace{-2em}where T$_{\rm S100}$ is the spin temperature in units of 100 K.

As a working hypothesis, we will assume that the HI seen in absorption
lies in the disk of NGC 1144, and exhibits similar kinematics to the
ionized gas. It can be seen from the 20 cm map of Condon et al. (Figure
11) that the brightest radio emission regions are the unresolved
nucleus (25.9 mJy) and the eastern extra-nuclear region (22.9 mJy:
angular size 2$\arcsec$ $\times$ 1$\arcsec$). The third extra-nuclear
region to the west of the nucleus is significantly fainter (9.4 mJy),
and probably does not contribute to any absorption profile.  Of the
strong emission regions, only the eastern radio source lies in
projection against the velocity field at 9000 km s$^{\rm -1}$. We will
adopt this source as a plausible continuum feature against which we
might see the higher velocity HI absorption feature of Figure
2.\footnote{Had we chosen the nuclear source, rather than the eastern
extra-nuclear source, the calculated optical depth would be similar
since its flux is almost identical} For an absorption feature seen
over one channel (43.8 km s$^{\rm -1}$) and a depth F$_{\rm abs}$ = 2
mJy, and F$_{\rm con}$ = 22.9 mJy, the neutral hydrogen column
density needed to create the absorption will be N$_{\rm HI}$ = 6.7
$\times$ 10$^{\rm 20}$ T$_{\rm S100}$ atoms cm$^{\rm-2}$ channel$^{\rm
-1}$. Since the feature is seen over three channels, the total column
density is similar to that seen commonly in the disks of galaxies in
emission, and is $\sim$2 $\times$ 10$^{\rm 21}$ T$_{\rm S100}$ atoms
cm$^{\rm -2}$. A single cloud covering the area of the eastern radio
emitting region, and with the above column density, would have a mass
of $\sim$8 $\times$ 10$^{\rm 6}$ M$_{\odot}$, the mass of a typical
giant molecular cloud (GMC). Hence our basic hypothesis that the
absorbing cloud lies in the disk of NGC 1144 is consistent with the
observations.

The next question to ask is why we don't see a large population of
such clouds in emission, as we apparently do in the NW part of the
disk?\footnote{We remind the reader that we do see some faint emission
centered on the nucleus over a few channels in the velocity range
9200-9300 km s$^{\rm -1}$.}  Here we are confronted with two possible
alternatives. The first hypothesis - which we call the ``minimal
absorption'' hypothesis - is that the clouds we see in absorption have
a small filling factor compared with the larger VLA beam (21$\arcsec$
$\times$ 17$\arcsec$).  How many such clouds could be hidden within
the C-array VLA beam before we would see the clouds in emission rather
than absorption? The 3$\sigma$ noise per channel was found to be
approximately 1 mJy beam$^{\rm -1}$, and so it would be possible to
hide 1.3 $\times$ 10$^{\rm 8}$ M$_{\odot}$ of HI within one beam and
still fail to detect it. If we assume that such emission would be in
the form of the clouds we detect in absorption, this implies an upper
limit of 16 absorber clouds per beam that could go undetected in
emission. As a point of reference, in the NW disk of NGC 1144, the
typical column density of gas seen in emission would imply
approximately 200 similar clouds per beam. This would argue for a
depletion in the total HI mass in that region of the disk, and would
be consistent with the scenario presented in $\S$6.1, in which some
process is destroying HI clouds in the SE part of the disk of NGC
1144.

A second hypothesis - we call the "extreme-absorption" hypothesis - is
that the entire HI profile from 8400 to 9000 km s$^{\rm -1}$ is
affected by {\it deep absorption} against the nucleus, which nullifies
any {\it emission} seen in the larger VLA C-array HI beam. We note that
the HI emission and absorption in the nucleus is spotty and not
``perfectly'' balanced. Such a scenario is strongly hinted at in
Figure 10, which shows the possibility of a broad absorption complex
in the nuclear region. 

Strong, broad (up to 700 km s$^{\rm -1}$ wide) HI absorption against
Seyfert nuclei is not uncommon (e.g. Dickey 1982,1986; Mirabel
1982,1983). Van Gorkom et al. (1989) detected HI in absorption (up to
$\sim$600 km s$^{\rm -1}$ wide) against the nuclei of radio elliptical
galaxies. These systems have absorption features which are consistent
with both infalling and outflowing HI gas. Further, IC 5063 has a 700
km s$^{\rm -1}$ HI absorption feature, with a depth of 10-15 mJy, seen
against its Seyfert 2 nucleus (Morganti, Oosterloo, \& Tsvetanov
1998). The absorption is seen blueward of the systemic velocity of IC
5063, indicating a blueshifted outflow of HI. The interpretation of
these various observations is that the HI is likely to be in the form
of an inflowing/outflowing stream, or complex of clouds, seen in
projection against a background continuum source (the AGN). If a high
column-density stream of HI were mixed, for example, with the ionized
gas associated with the velocity discontinuity seen in the nucleus
(Figure 11), then such absorption would have a profound effect on the
HI emission spectrum.

To explore the feasibility of the extreme-absorption hypothesis, we
will assume that, in the absence of absorption, NGC 1144 would have a
normal double-horned HI profile with a velocity width similar to the
observed spread in the ionized gas velocities (1100 km s$^{\rm -1}$)
and a single-channel flux of 10 mJy over that mid-range of the
spectrum. We can then ask what column density of HI would be required
to be seen throughout the range of 8400-9000 km s$^{\rm -1}$ (the
velocity jump seen near the nucleus in the ionized gas) to reduce the
putative HI emission line profile to zero flux?  If F$_{\rm abs}$ = 10
mJy and the continuum source is the nucleus (F$_{\rm con}$ = 25.9 mJy)
then we derive N$_{\rm HI}$ = 4.2 $\times$ 10$^{\rm 22}$ T$_{\rm
S100}$ atoms cm$^{\rm -2}$ over the full 600 km s$^{\rm -1}$ range
seen at the velocity discontinuity. This must be considered an upper
limit to the true column density since the radio source is not
resolved in the observations made by Condon et al. (1990). If we
assume it is just resolved, the implied cloud mass of the HI absorbers
would be 1.6 $\times$ 10$^{\rm 8}$ M$_{\odot}$, assuming that the area
of the nuclear region is 1.65 square arcseconds, the approximate area
of the beam from Condon et al. (1990).
 
The advantage of the extreme-absorber hypothesis over the
minimal-absorber picture, is that it can readily explain the lack of
HI in emission from the mid- to SE disk of NGC 1144, and indeed from
the HI data-cube over the entire range of velocities in excess of 8661
km s$^{\rm -1}$. It also is very testable. Higher resolution HI
observations should separate any, so far only hypothetical, deep
absorption in the nucleus from extended HI emission in the disk.

\section{HI Distribution and Kinematics of a "Ultra-violet Excess" Dwarf 
Companion in the Arp 118 Group = KUG 0253-003}

We have discovered HI emission from a position ($\sim$8$\arcmin$ to
the NE of Arp 118 = 256 kpc) corresponding to the same spatial
location as a galaxy found by Takase \& Miyauchi (1988) in a survey of
ultraviolet-excess galaxies (KUG 0253-003 or PGC 011066).
NED\footnote{The NASA/IPAC Extragalactic Database (NED) is operated by
the Jet Propulsion Laboratory, California Institute of Technology,
under contract with the National Aeronautics and Space
Administration.} lists this galaxy as a ``spiral'' with a blue
magnitude of 16.5, and with major \& minor axes of 0$\farcm$3 $\times$
0$\farcm$3 and no cataloged redshift. There are no obvious spiral arms
in the DSS image, and it appears as a high-surface brightness, compact
galaxy.

Figure 13 contains the global HI profile of the companion, which is
narrow and single peaked, with a $\Delta$V$_{\scriptsize \onehalf}$ of
51 km s$^{\rm -1}$. The systemic velocity (V$_{\rm HI}$) is 8749 km
s$^{\rm -1}$, which gives a velocity difference of $\sim$460 km
s$^{\rm -1}$ between the distant, uv-excess dwarf companion and Arp
118. Assuming it to be part of the extended Arp 118 group, the HI mass
of KUG 0253-003 is 3.4 $\times$ 10$^{\rm 9}$ M$_{\odot}$, roughly half
the value of Arp 118 (see Table 1). This ``dwarf'' galaxy therefore
appears to be somewhat rich in atomic hydrogen. In Figure 14 we
present the integrated HI image of the companion, overlayed with a DSS
grey-scale image. The HI emission is smooth, with a peak that appears
to be somewhat off-set from the bright centrally concentrated light
distribution of KUG 0253-003. The HI emission appears to be elongated
to the SW (towards Arp 118).

Figure 15 contains the 3 channel maps with HI emission from the
companion. Figure 16 shows the mean velocity field of KUG 0253-003,
which, though we barely resolve the galaxy, hints at rotation, with a
major kinematic axis that runs approximately SE to the NW.

One might speculate that the dwarf is undergoing a major episode of
star formation, perhaps as a result of an interaction with Arp
118. The difference in velocity (460 km s$^{\rm -1}$) and projected
separation suggest a time-scale of 5 $\times$ 10$^{\rm 8}$ years to
get to its present location if it did pass close to Arp 118 in the
past. This is the typical crossing-time for a small group and does not
seem unreasonable.

\section{Conclusions}

A mapping of the neutral and ionized gas in the Arp 118 system has
revealed the following:

1) The HI emission, in addition to being highly disturbed, is
distributed non-uniformly throughout the disk of NGC 1144, being
highly concentrated in the NW region of the ring away from the nuclear
star formation complexes and the Seyfert 2 nucleus. This distribution
is anti-correlated with the most powerful CO emission which lies in
the SE part of the disk. Ionized gas is seen distributed throughout
the entire disk of NGC 1144. No HI or H$\alpha$ emission is seen
associated with the elliptical companion NGC 1143.

2) Unlike the huge spread in the ionized gas velocities in the disk of
NGC 1144 (1100 km s$^{\rm -1}$), strong HI emission is observed only over
one-third of this interval and is consistent with HI emission
associated with the NW part of the ionized disk. HI emission over a
600 km s$^{\rm -1}$ interval (which covers most of the mid-disk and SE
half of the ionized disk) is missing from the velocity channels.

3) Observations of the ionized gas component of the disk of NGC 1144
show that, in addition to the large-scale rotation of the galaxy over
the range of 1100 km s$^{\rm -1}$, there exists a sudden jump in
velocity of the ionized gas in the nuclear regions of the galaxy of
approximately 600 km s$^{\rm -1}$. If we adopt the systemic velocity
of the H$\alpha$ disk to be 9000 km s$^{\rm -1}$ determined from the
large-scale velocity field away from the nucleus, then this implies
blue-shifted gas streaming motions within the nucleus. Alternatively,
the observations may imply that the nucleus itself is moving at a
velocity of 600 km s$^{\rm -1}$ with respect to the host
disk. However, CaII H \& K absorption lines support a nuclear velocity
of around 9000 km s$^{\rm -1}$ suggesting that it is the ionized gas
that has the peculiar motion in the nucleus.

4) Very weak HI absorption is seen against the radio continuum sources
(including an off-center nucleus) which are concentrated in the SE
section of the ring.  We interpret the absorption lines in two ways:
a) the ``minimal-absorption'' hypothesis, in which the weak absorption
lines are assumed to be the only absorption present, and b) the
``extreme-absorption'' hypothesis, in which it is assumed that there is
strong and deep nuclear HI absorption over the same velocity range as
the H$\alpha$ velocity discontinuity (600 km s$^{\rm -1}$). Hypothesis
a) results in an explanation for the missing HI in the mid-to-SE disk
of NGC 1144 as a depletion of HI clouds by at least a factor of 10
over the NW disk. However, b) offers the advantage that it explains
the missing HI in the disk of NGC 1144 as being due to a high
column-density stream of HI seen in absorption against the nucleus,
negating the HI emission in the larger-scale disk. If true, b)
suggests a streaming of high column-density HI gas (N$_{\rm H}$ $>$ 4
$\times$ 10$^{\rm 22}$ atoms cm$^{\rm -2}$) is mixed with ionized gas
and is emanating from the nucleus at up to 600 km s$^{\rm -1}$. The
total HI mass of such a stream could be as high as 1.8 $\times$
10$^{\rm 8}$ M$_{\odot}$ in the neutral hydrogen component alone.

5) We detect an HI-rich dwarf galaxy 8$\arcmin$ to the NE of Arp
118. The dwarf galaxy, KUG 0253-003, was detected in a survey of ``uv
excess'' galaxies, and shows a disturbed HI distribution with an
extension pointing towards Arp 118.  It is possible that the galaxy
has interacted with the Arp 118 system in the past (500 Myr ago, based
on its projected separation and velocity difference).

It is clear that high signal-to-noise HI observations of much higher
spatial resolution are required to determine the true nature of the
absorption spectrum of NGC 1144. The source that is naturally
implicated is the nuclear source, which we know has a flux of about 25
mJy at $\lambda$20 cm. Although observations on the scale of several
arcseconds will be required to determine whether the disk of NGC 1144
is truly asymmetric in its HI content, the study of the postulated
high-column density stream will require sub-arcsecond observations of
the absorption lines over the nuclear source. If we are correct in our
extreme-absorber hypothesis, the absorption lines could be as deep as
10 mJy (against a source of 25 mJy) over a velocity range from 8400 to
9000 km s$^{\rm -1}$.

Acknowledgements

The authors would like to thank Yu Gao (U. of Toronto) for helpful
exchanges of information about his CO observations of Arp
118. Particular thanks are due to Evan Skillman (U. of Minnesota)
whose insight has been of crucial importance. We also thank Matt Malkan for 
providing us with the HST image of NGC 1144. This work is supported
by NSF grant AST-9319596.

%
%

%
%

\clearpage
\begin{figure}
\figurenum{1}
\figcaption[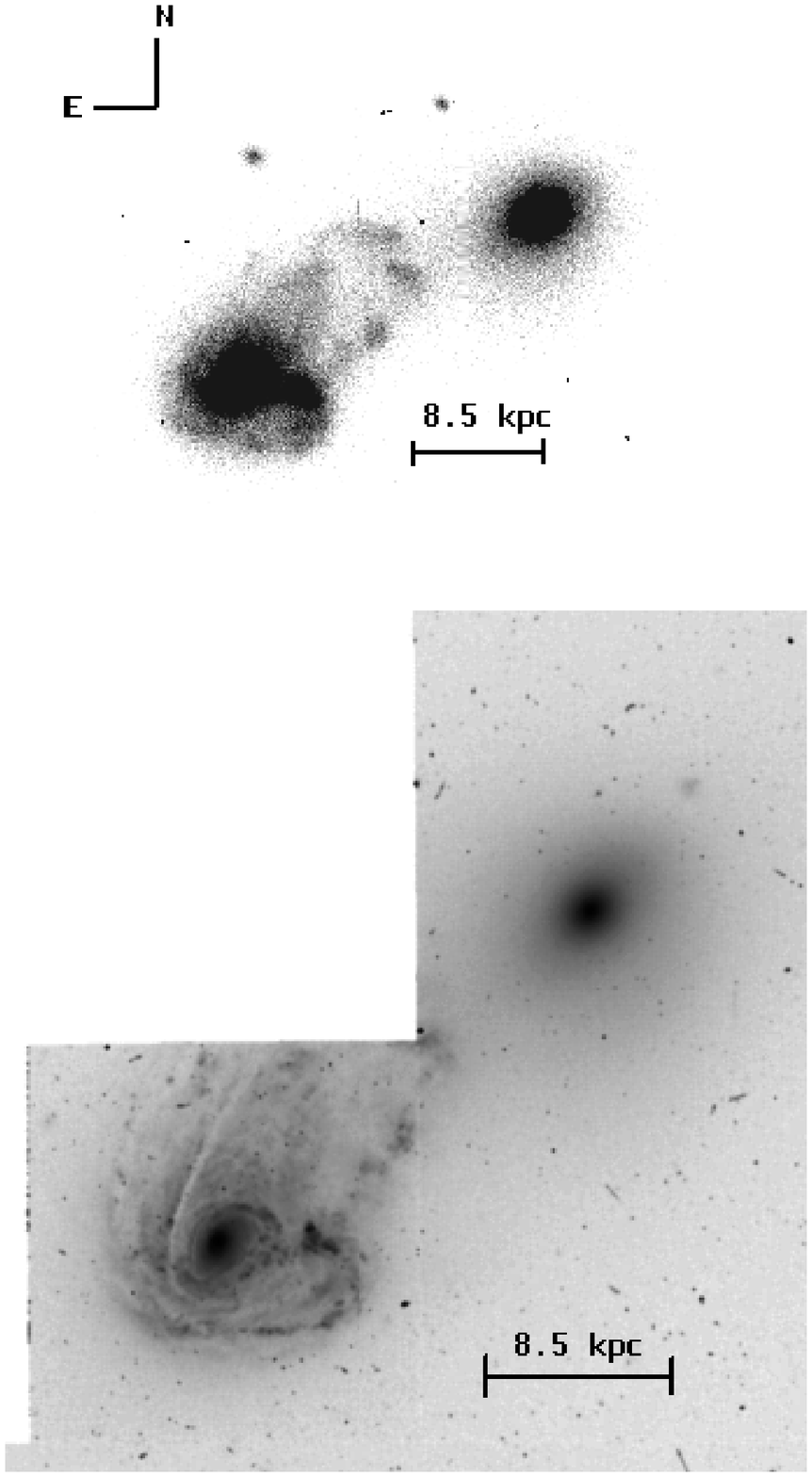]{H$\alpha$ image (top), taken with Steward 
Observatory 90-inch telescope, and HST F606W image (bottom), courtesy
of Matt Malkan. The H$\alpha$ image courtesy of M. A. Bransford and
A. P. Marston from unpublished data}
\end{figure}

\begin{figure}
\figurenum{2}
\caption[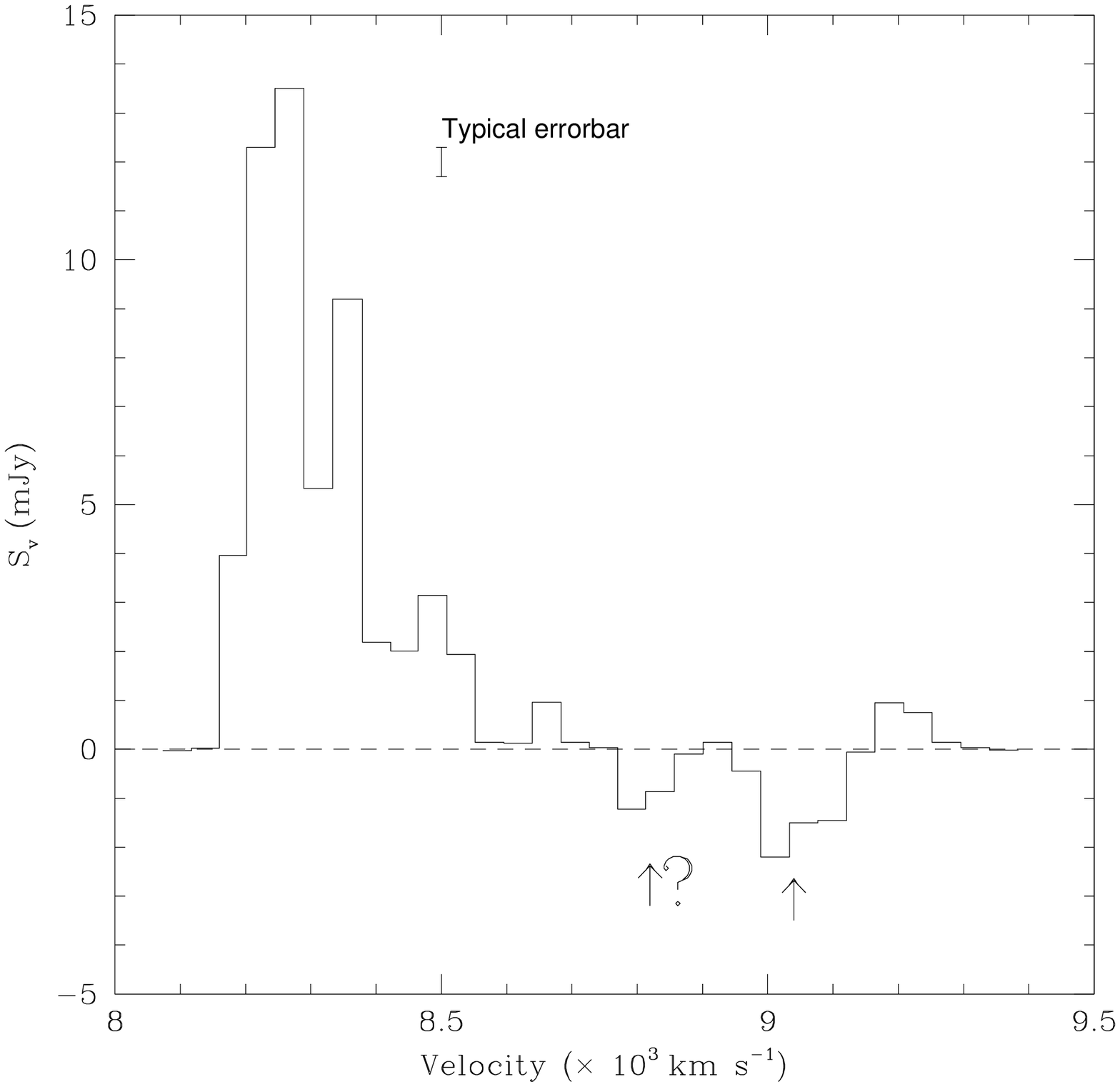]{Global HI profile of Arp 118. The arrows point to 
two absorption features. The fainter absorption feature at $\sim$8800 km 
s$^{\rm -1}$ is of marginal significance and further observations will be 
necessary to confirm its reality.}
\end{figure}

\begin{figure}
\figurenum{3}
\caption[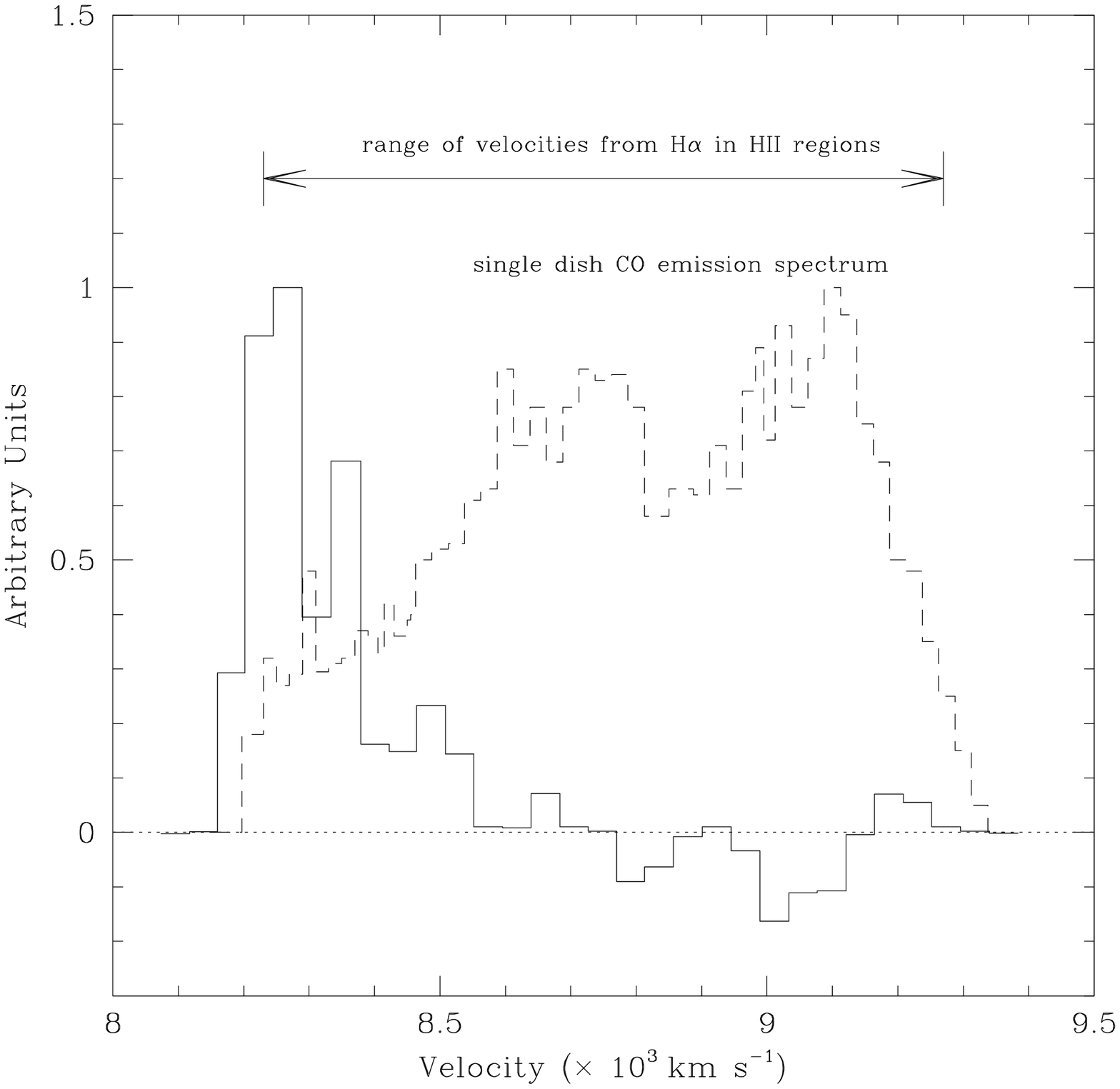]{Global HI profile of Arp 118 (solid line) 
over-plotted by global profile of CO emission (dashed line, GSDR). The double 
headed arrow indicates the range of velocities of the H~II regions around the 
ring (H89). Note how the HI emission peaks at low velocities, whereas the CO 
emission peaks at high velocities. The peaks of the HI and CO profiles
were arbitrarily normalized to a value of 1.0.}
\end{figure}

\begin{figure}
\figurenum{4}
\caption[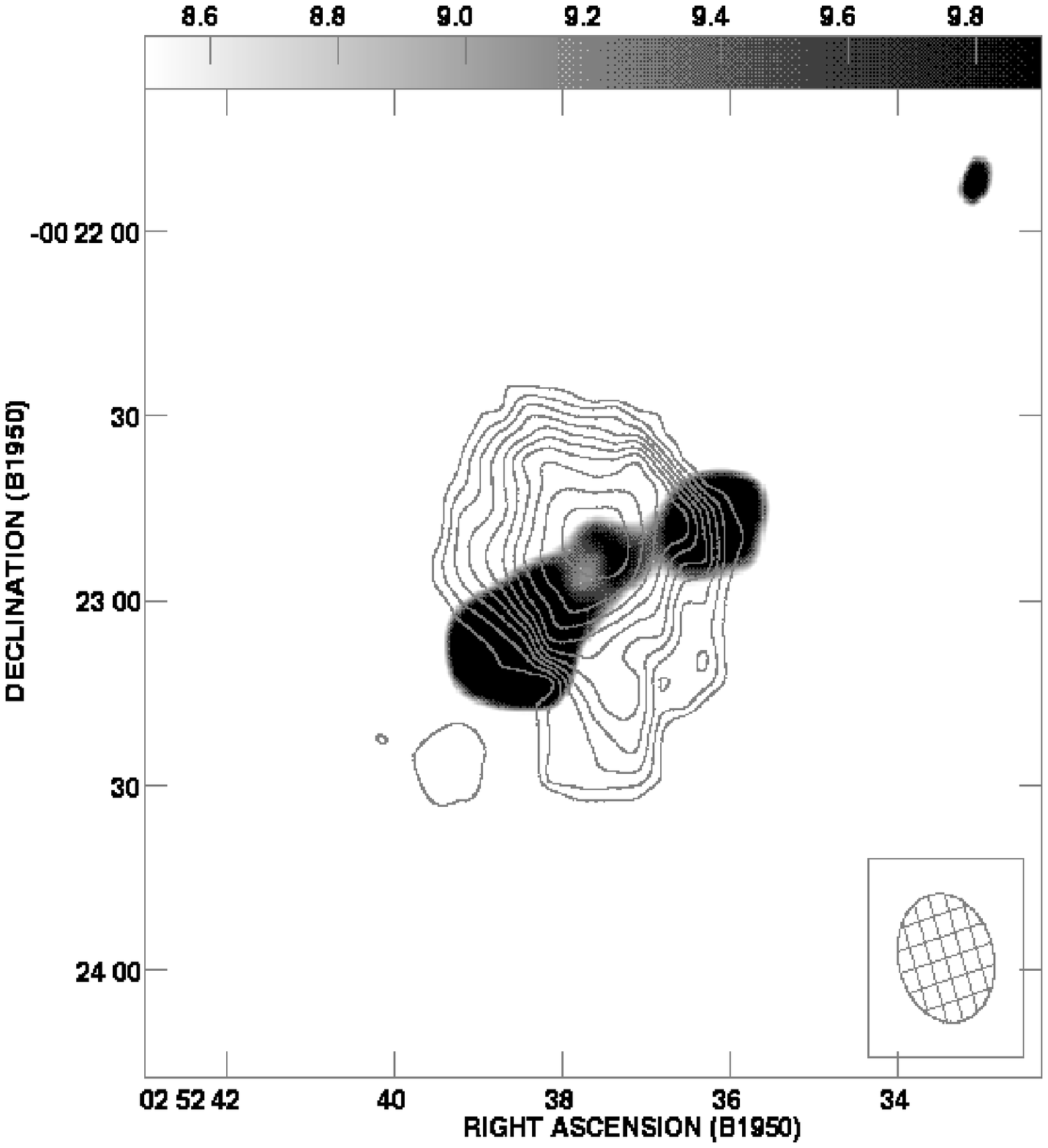]{Grey scale Digital Sky Survey image of Arp 118 
with a contour map of the integrated HI distribution. The contour increment 
is 50.0 Jy beam$^{\rm -1}$ m s$^{\rm -1}$, and the level of the lowest contour 
is 100 Jy beam$^{\rm -1}$ m s$^{\rm -1}$.}
\end{figure}

\begin{figure}
\figurenum{5}
\caption[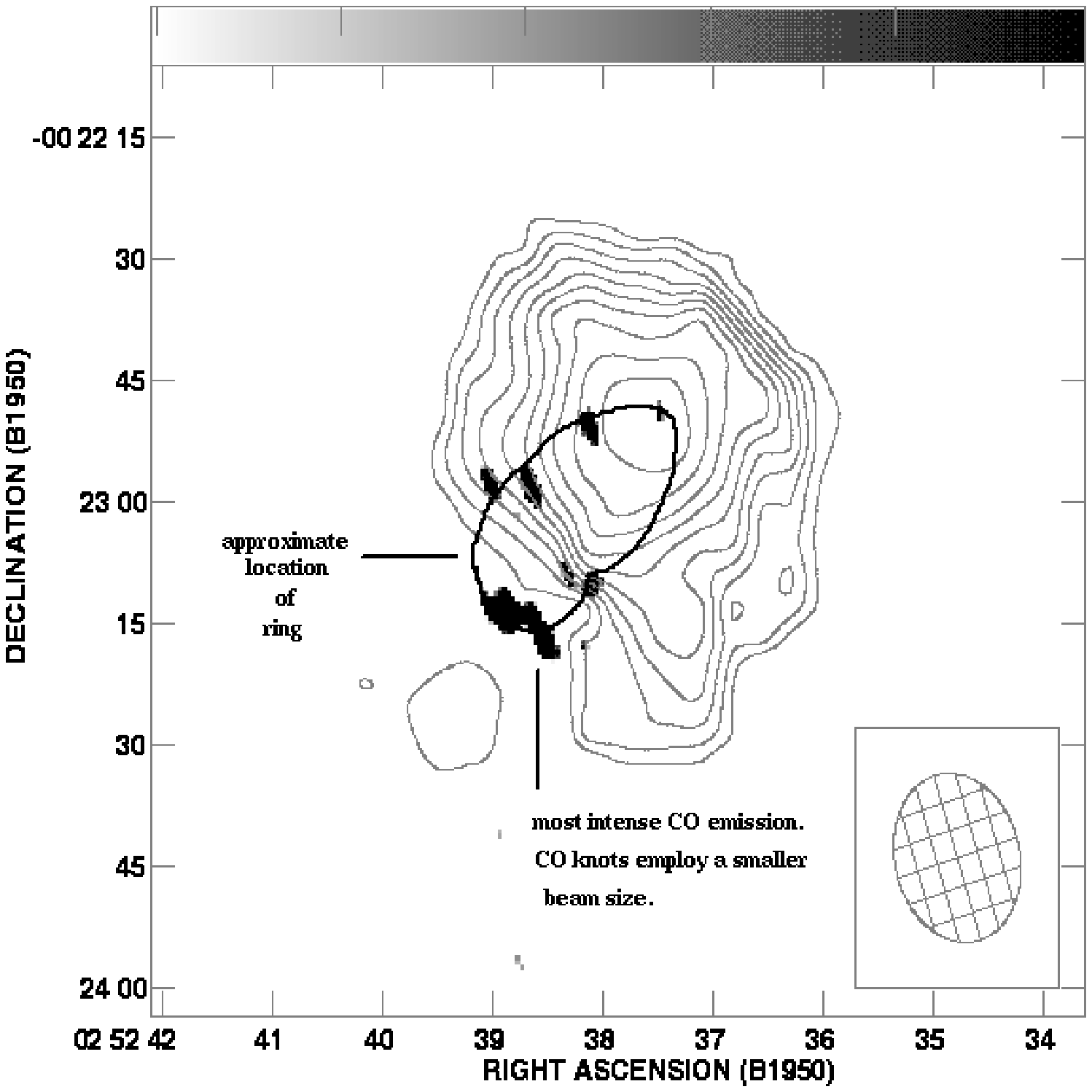]{Integrated HI contour map overlayed with a 
grey-scale image of the CO knots imaged by GSDR using the Plateau de Bure 
Interferometer. The CO knots were detected at a  different resolution 
(5$\farcs$3 $\times$ 2$\farcs$5 beam) than the VLA beam.}
\end{figure}

\begin{figure}
\figurenum{6}
\caption[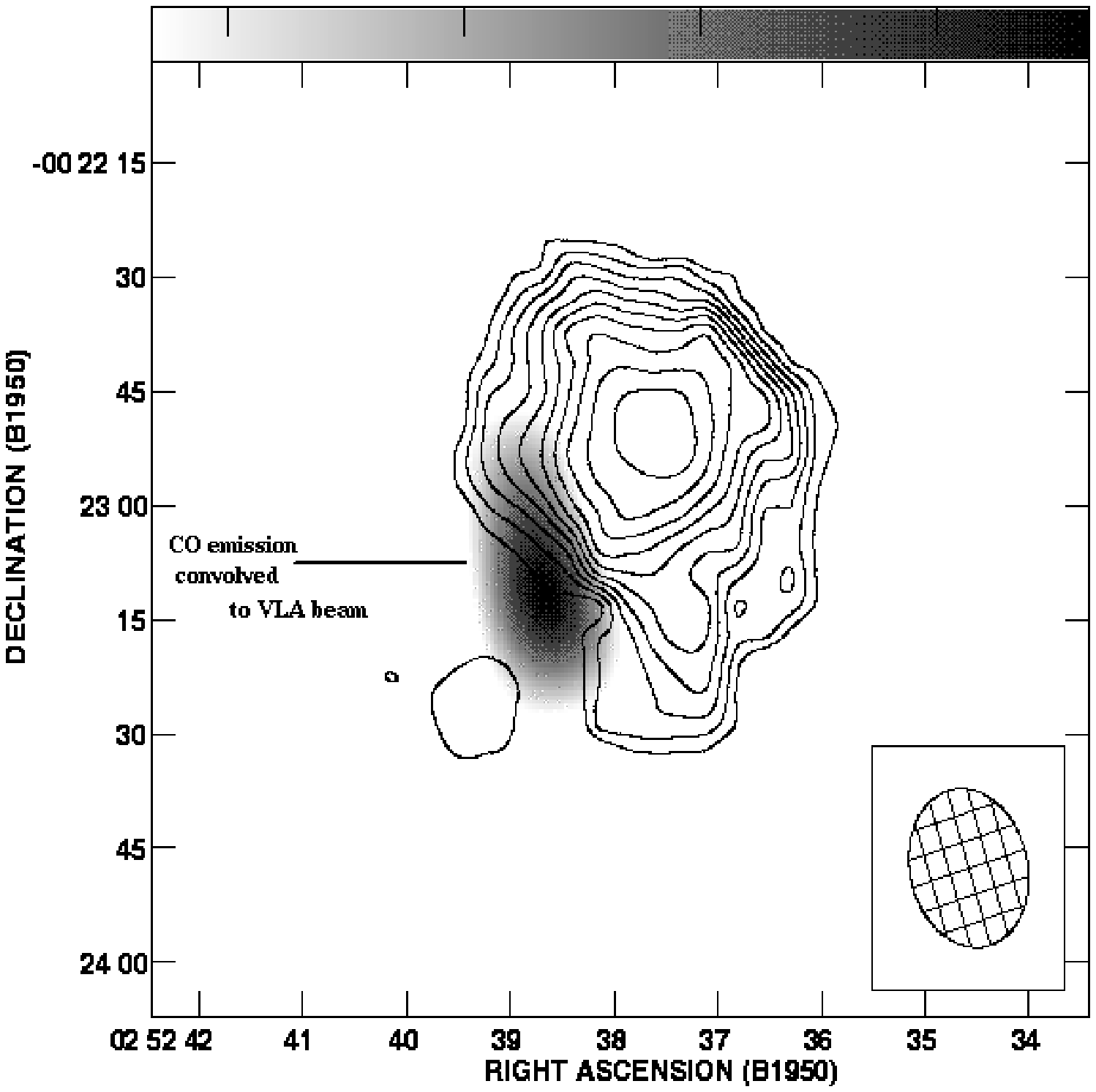]{Same as Figure 6, except the CO emission is 
convolved to the same VLA beam shape used in the HI observations.}
\end{figure}

\begin{figure}
\figurenum{7}
\caption[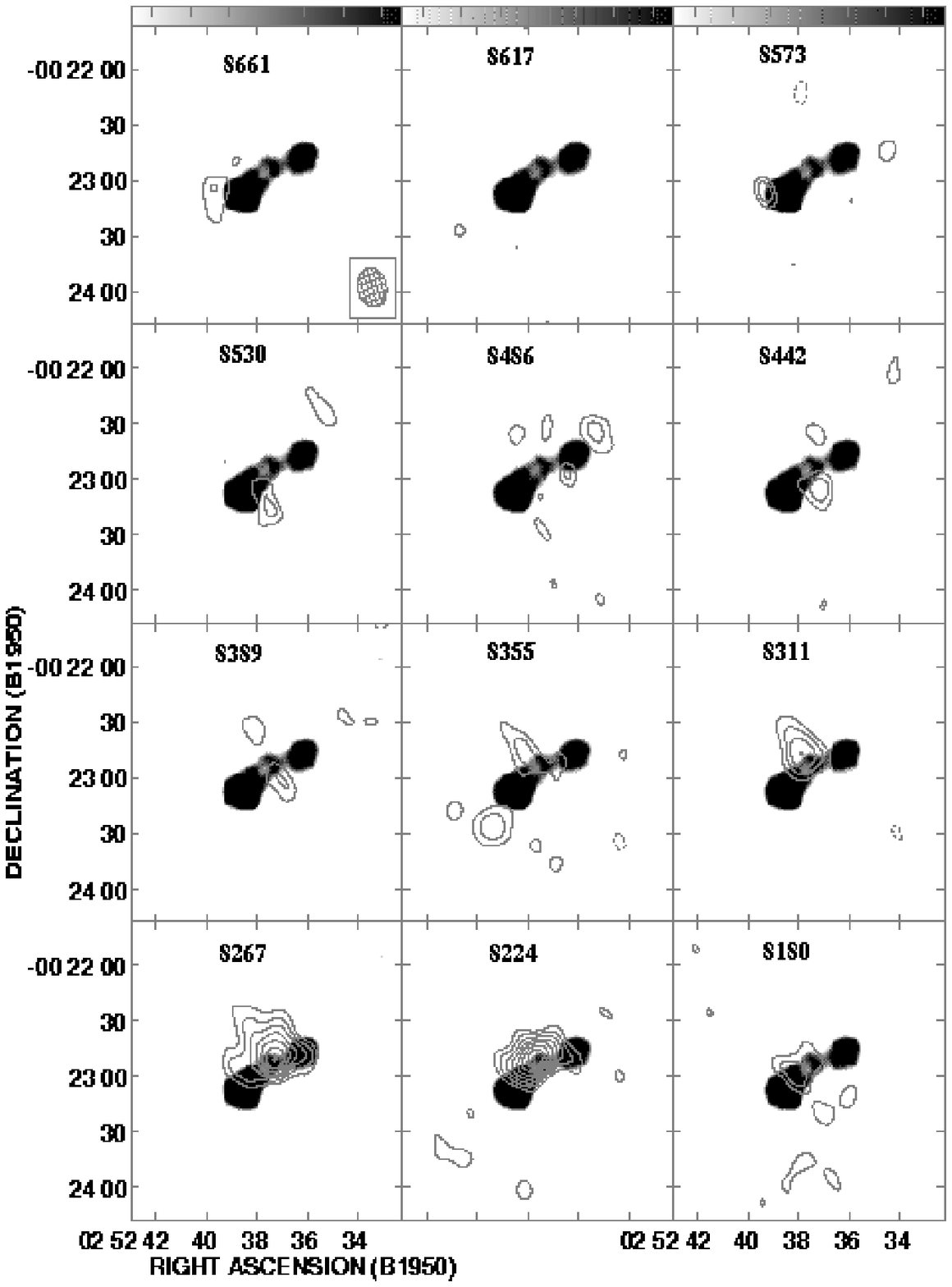]{Contour plots of the 12 channel maps which contain 
detectable HI emission, overlayed with the DSS gray-scale image of Arp 118. The
velocity (in km s$^{\rm -1}$) is displayed in the upper right. The channel 
width is 43.8 km s$^{\rm -1}$ in the rest frame of the galaxy. The contour 
increment is 3.6 $\times$ 10$^{\rm -4}$ Jy beam$^{\rm -1}$, and the lowest 
contour displayed is at 3$\sigma$.}
\end{figure}

\begin{figure}
\figurenum{8}
\caption[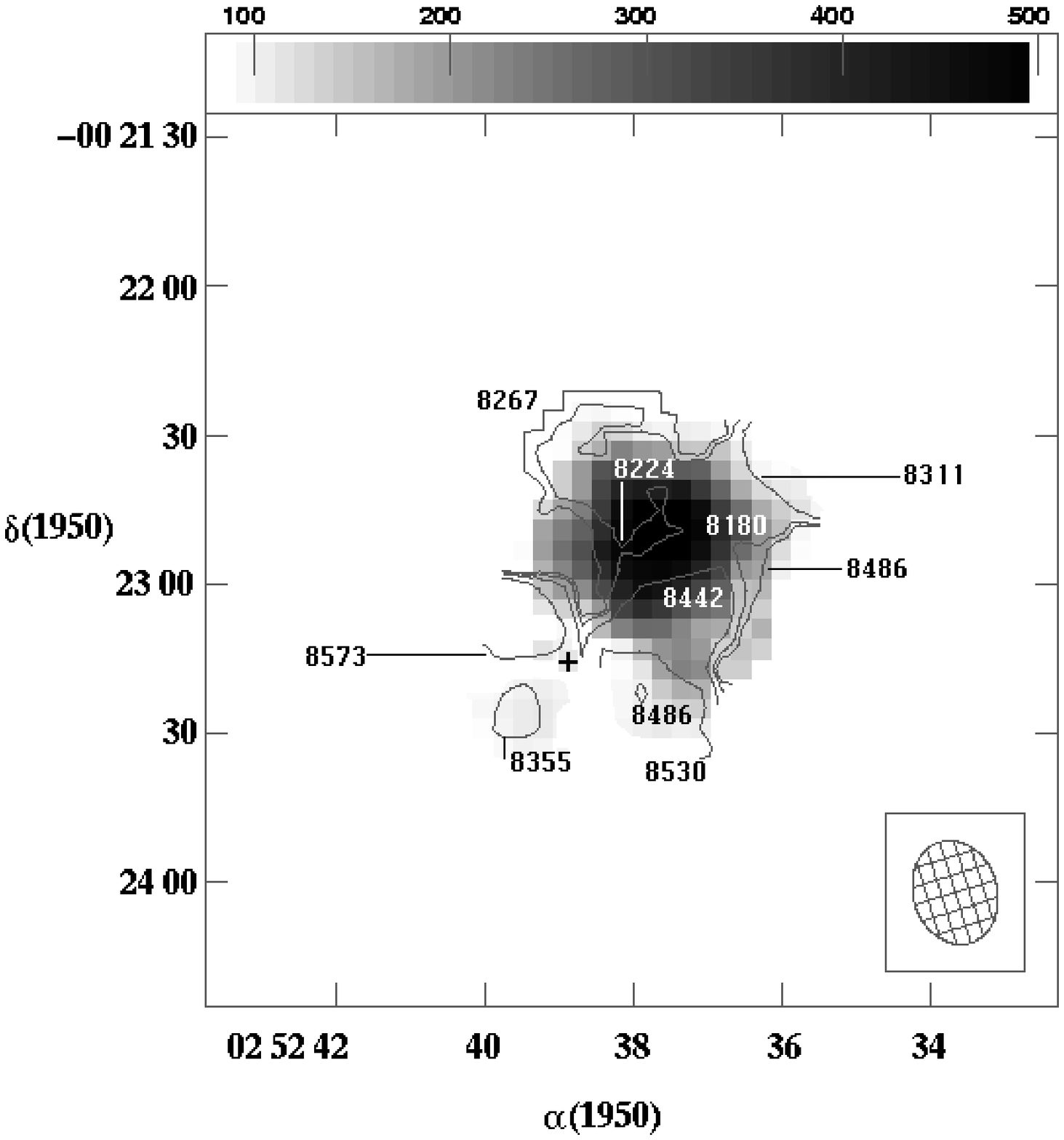]{Mean velocity field of Arp 118, with the velocity 
contours labeled in units of km s$^{\rm -1}$. The cross denotes the
location of the nucleus of NGC 1144.}
\end{figure}

\begin{figure}
\figurenum{9}
\caption[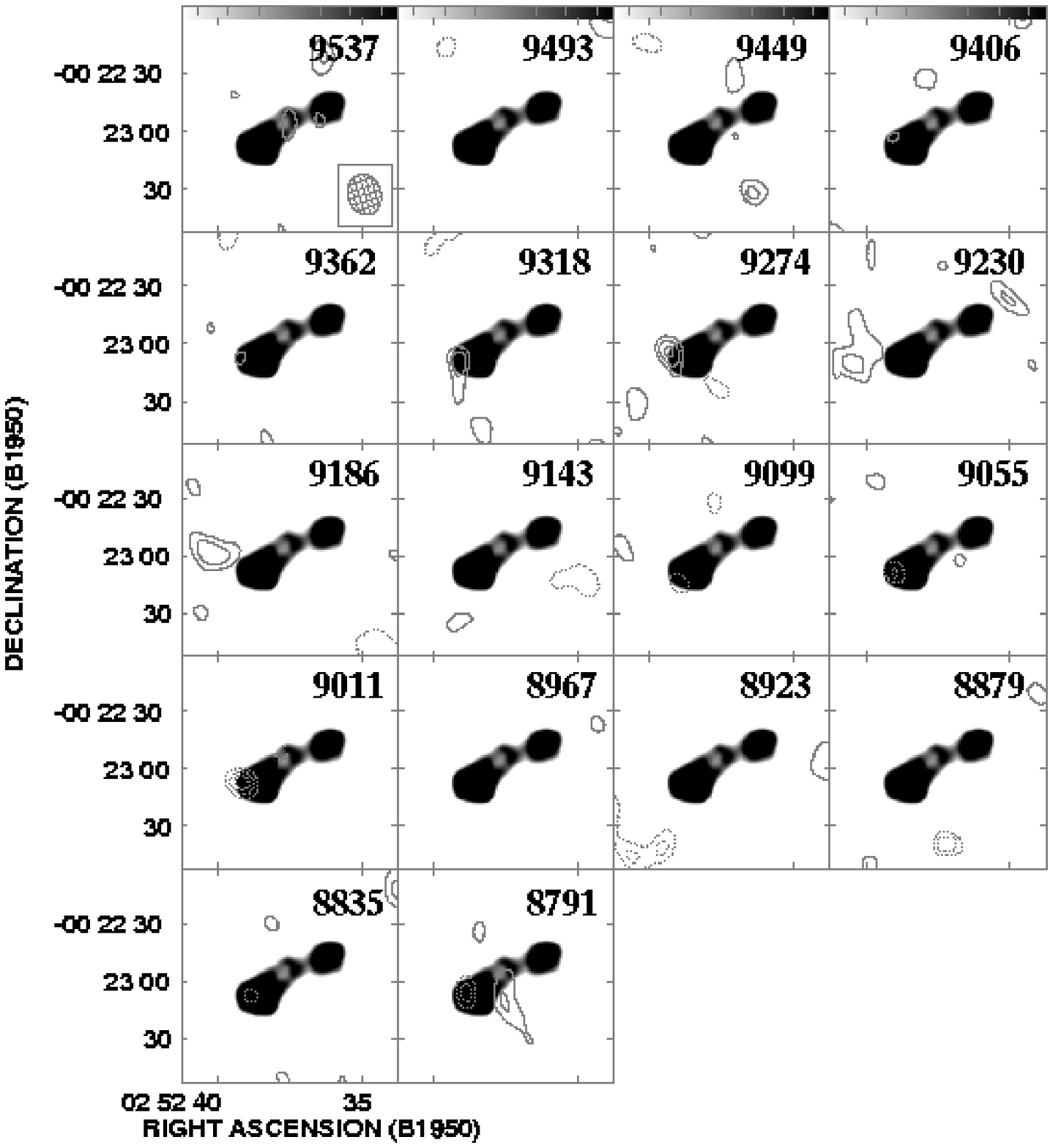]{A sequence of channel maps containing the faint 
absorption and emission features discussed in the text. The contour
increment is $\pm$3.3 $\times$ 10$^{\rm -4}$ Jy beam$^{\rm -1}$ and
the lowest contour displayed is 3$\sigma$. The dotted grey lines
denote negative contours and the solid grey lines denote
emission. Note the absorption features centered on the nucleus most
strongly in the 8791 and 9011 km s$^{\rm -1}$ channels, and the
emission features centered near the nucleus in velocity range 9186 to
9318 km s$^{\rm -1}$}
\end{figure}

\begin{figure}
\figurenum{10}
\caption[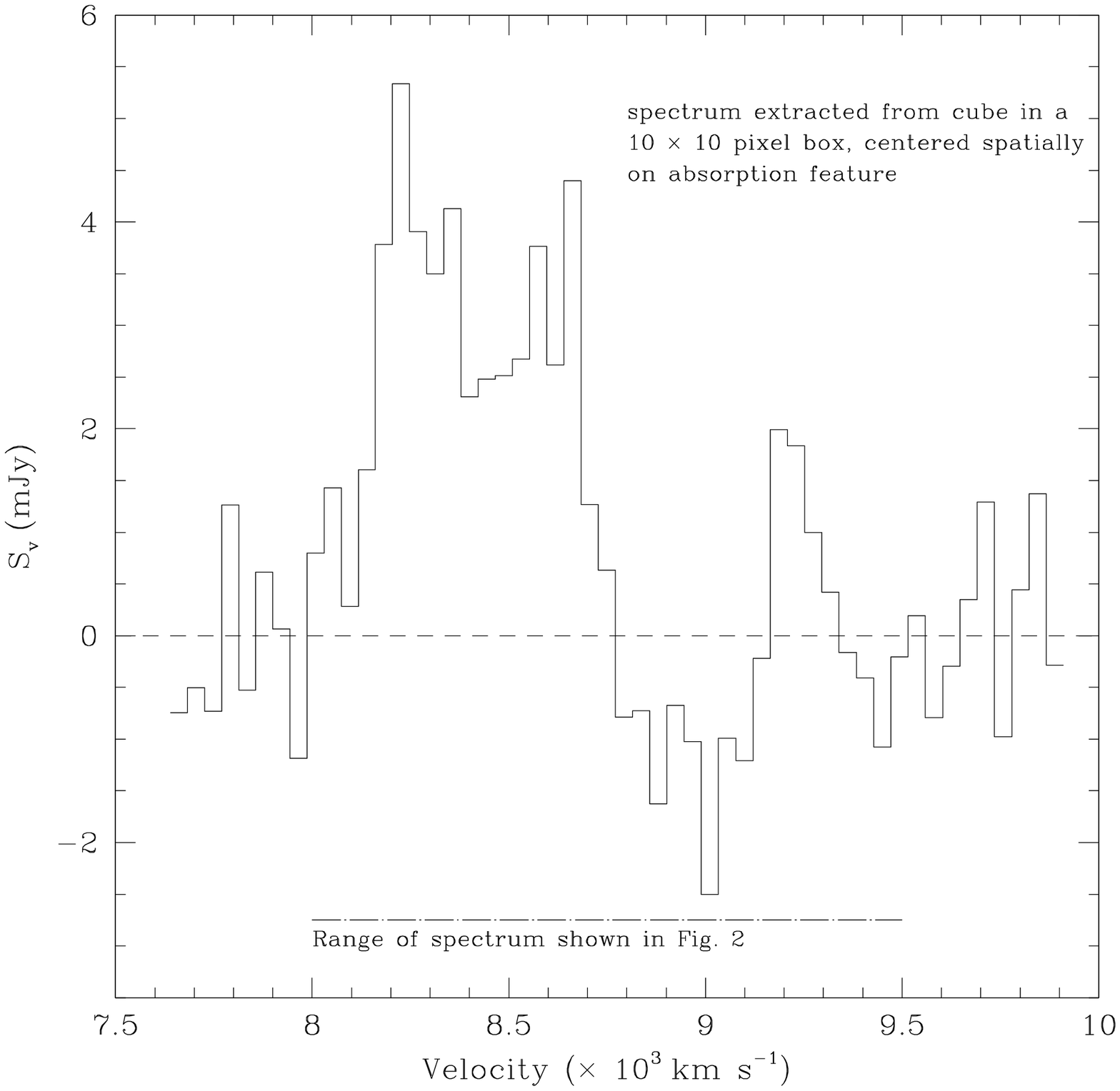]{HI profile, extracted from data cube in 
a 10 $\times$ 10 pixel box centered on the nuclear absorption
feature. This profile is the integrated emission and absorption
associated with the nuclear source only, and therefore differs
significantly from the global profile of Figure 2.}
\end{figure}

\begin{figure}
\figurenum{11}
\caption[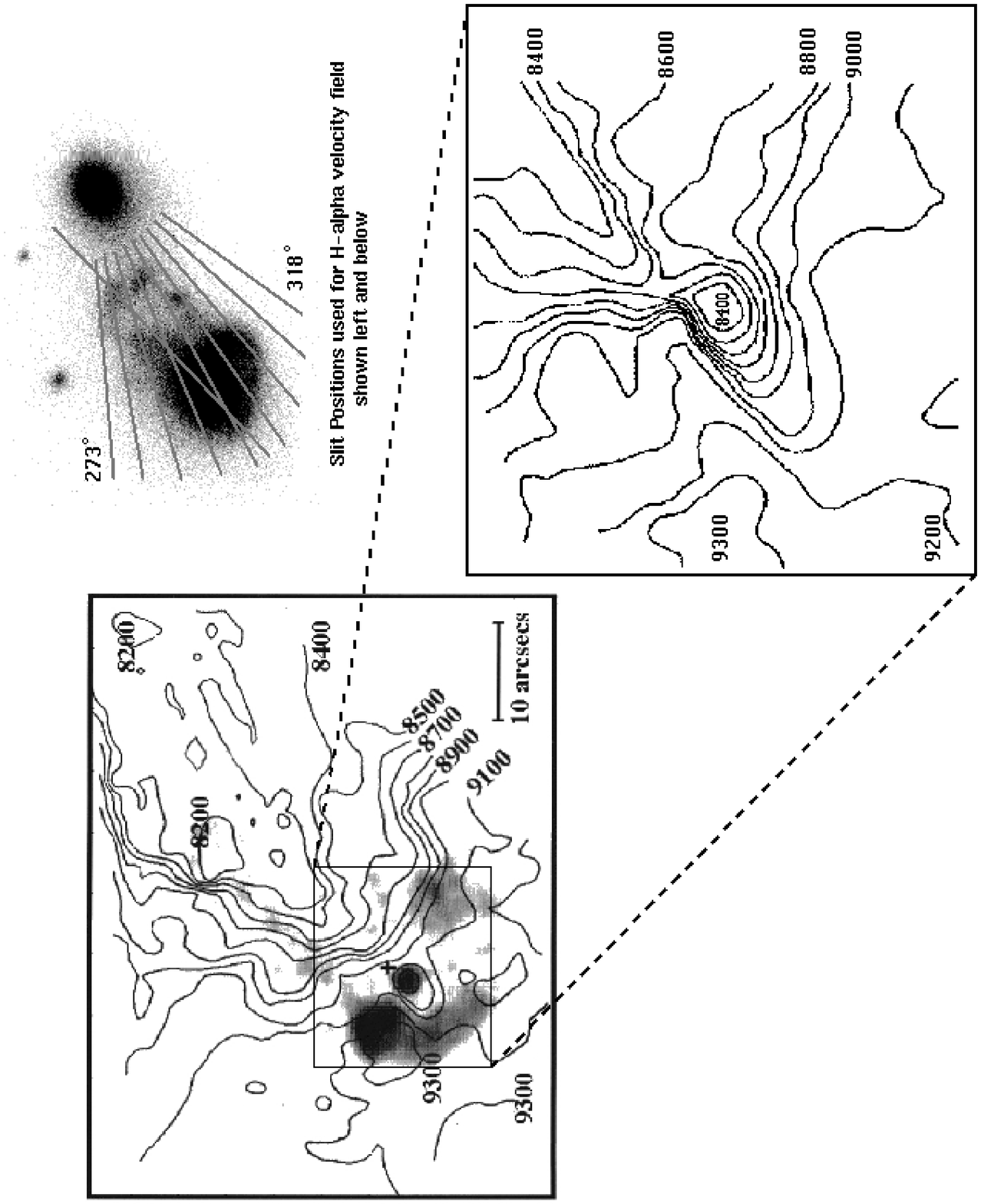]{Contour plot of the H$\alpha$ velocity field of 
the disk of Arp 118 (upper left). In the upper right are the slit positions 
used for determining the H$\alpha$ velocity field, and in the lower right an 
inset shows more detail of the kinematics of the nuclear region. Note the 
velocity discontinuity in the nuclear region of 600 km s$^{\rm -1}$. Finally, 
in the upper left, we overlay a grey-scale representation of the high 
resolution $\lambda$20 cm image of NGC 1144 from Condon et al. (1990). In 
making the overlay between the 20 cm continuum image and the velocity field, we
have assumed that the position of the velocity discontinuity is the nucleus.}
\end{figure}

\begin{figure} 
\figurenum{12}
\caption[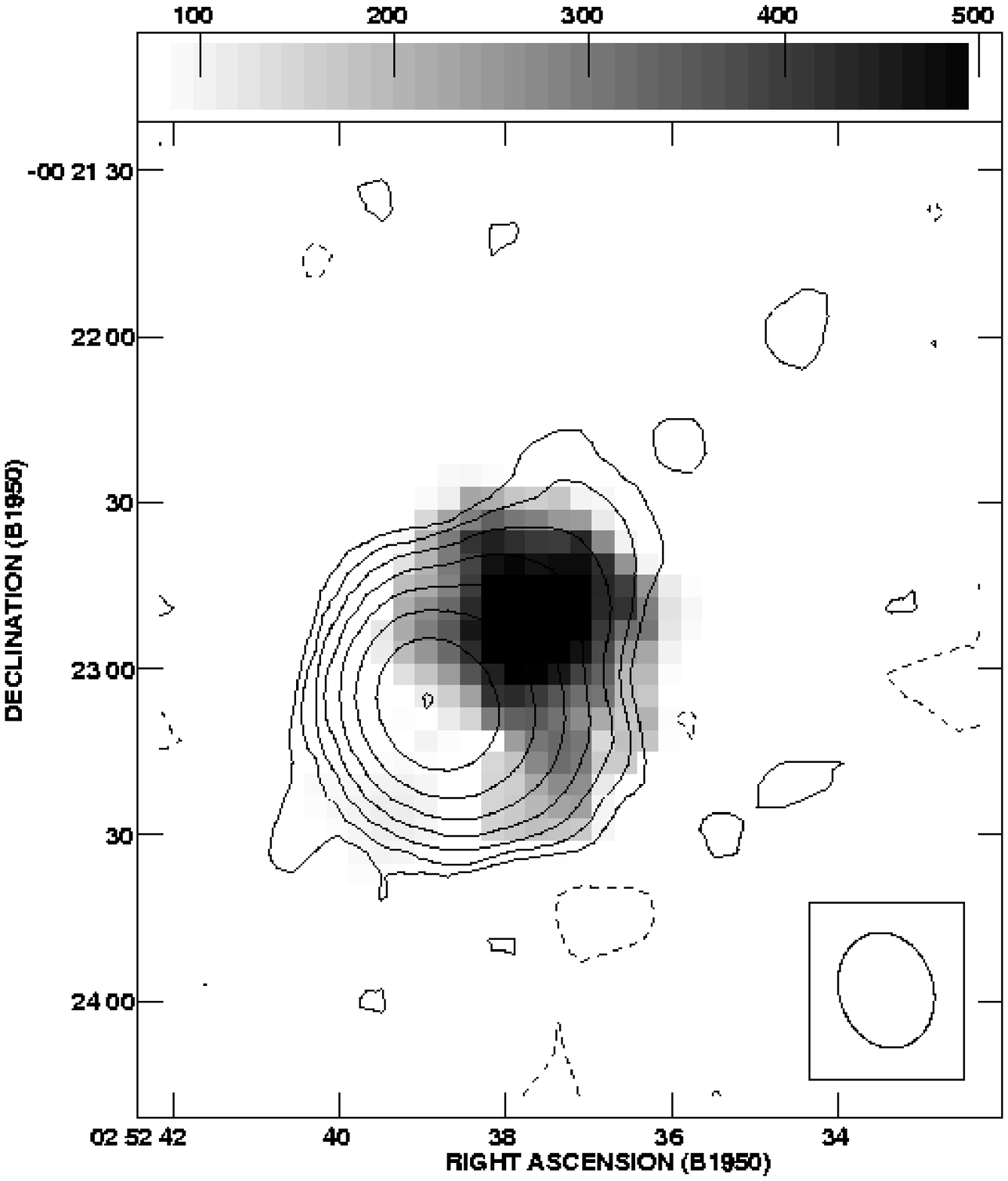]{Contour plot of the continuum emission overlayed 
on a greyscale image of the integrated HI emission, from Arp 118. The
contours are (-2,2,4,8,16,32,64,128,256) $\times$ 3.6 $\times$
10$^{\rm -4}$ Jy beam$^{\rm -1}$. The grey scale flux ranges from 80
to 500 Jy beam$^{\rm -1}$ m s$^{\rm -1}$.}
\end{figure} 

\begin{figure}
\figurenum{13}
\caption[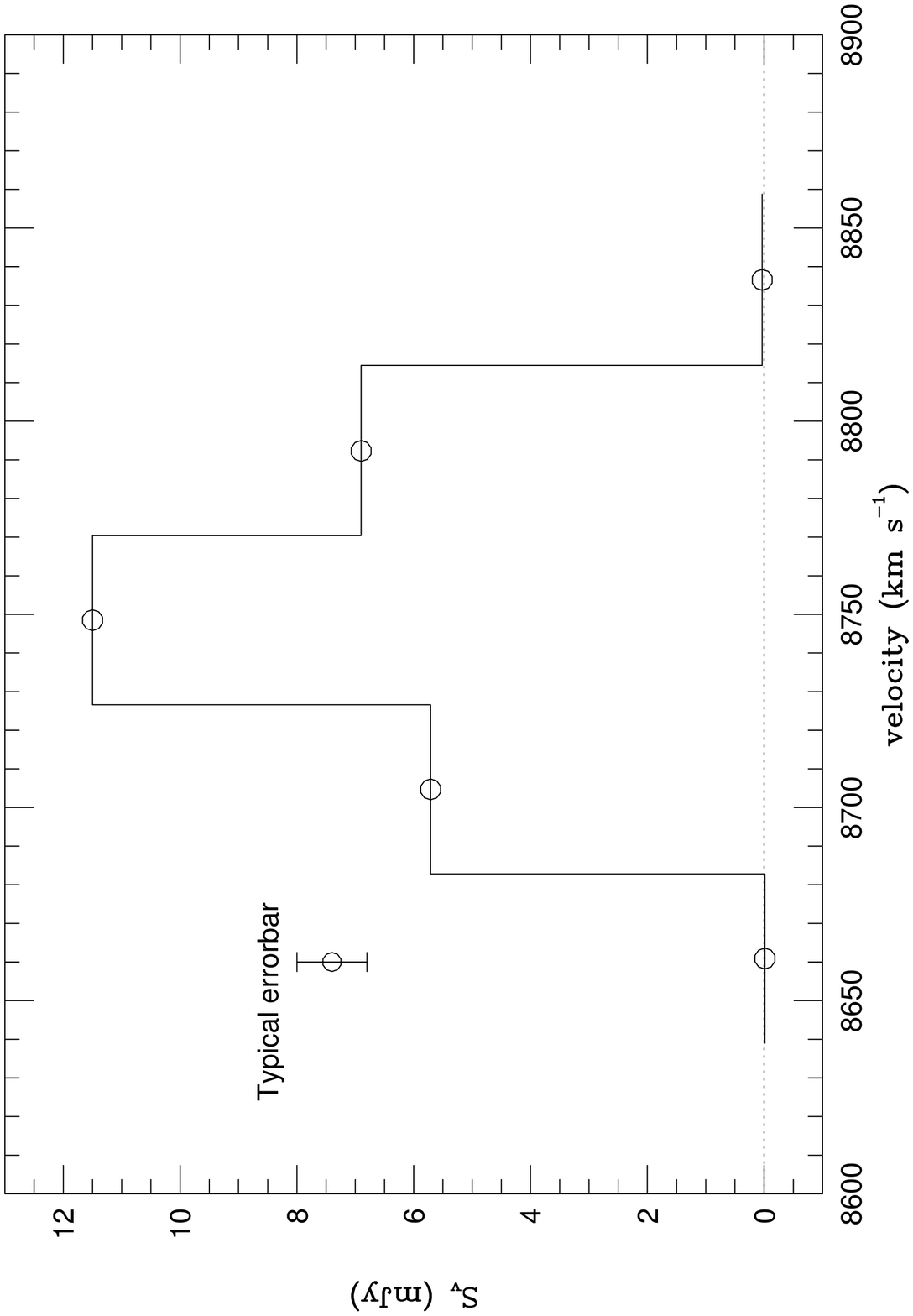]{Global HI profile of dwarf companion.}
\end{figure}

\begin{figure}
\figurenum{14}
\caption[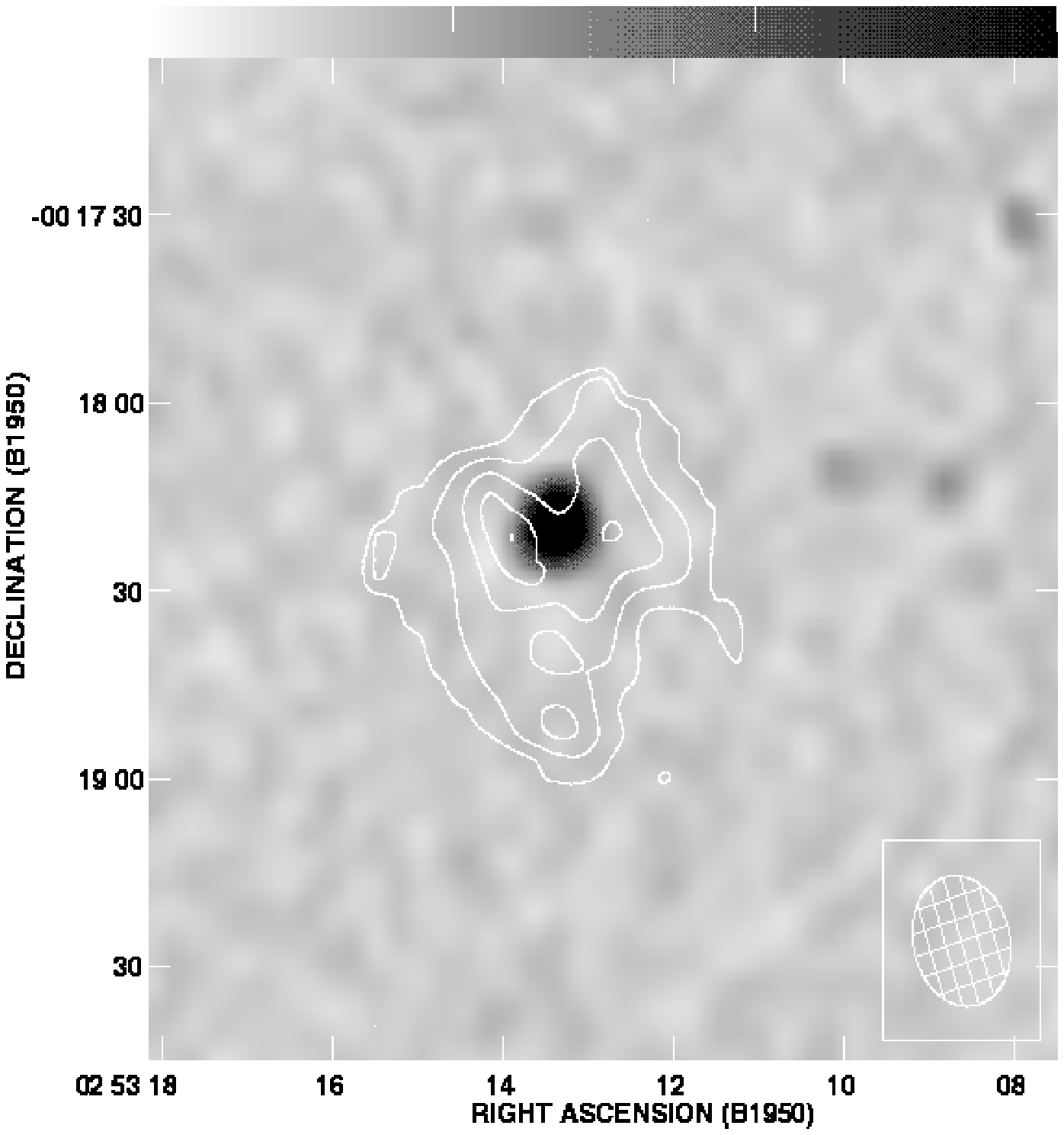]{Grey scale DSS image of dwarf companion with a 
contour map of the integrated HI distribution. The contour increment is 25 Jy 
beam$^{\rm -1}$ m s$^{\rm -1}$, and the level of the lowest contour is 75 Jy 
beam$^{\rm -1}$ m s$^{\rm -1}$.}
\end{figure}

\begin{figure}
\figurenum{15}
\caption[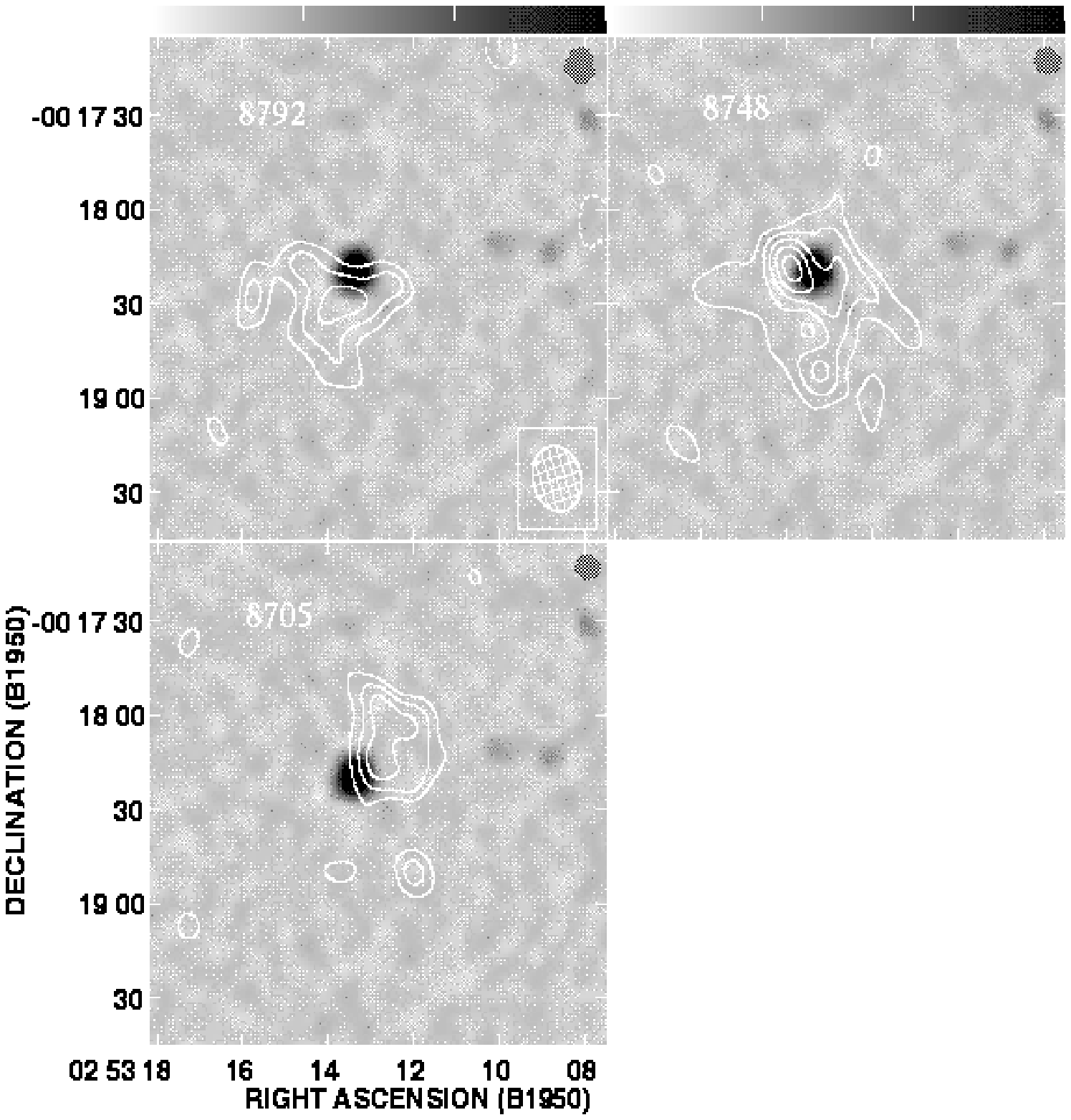]{Channel maps containing HI emission from dwarf 
companion. The contour increment is 3.3 $\times$ 10$^{\rm -4}$ Jy beam$^{\rm 
-1}$ and the lowest contour displayed is 3 $\sigma$.}
\end{figure}

\begin{figure}
\figurenum{16} 
\caption[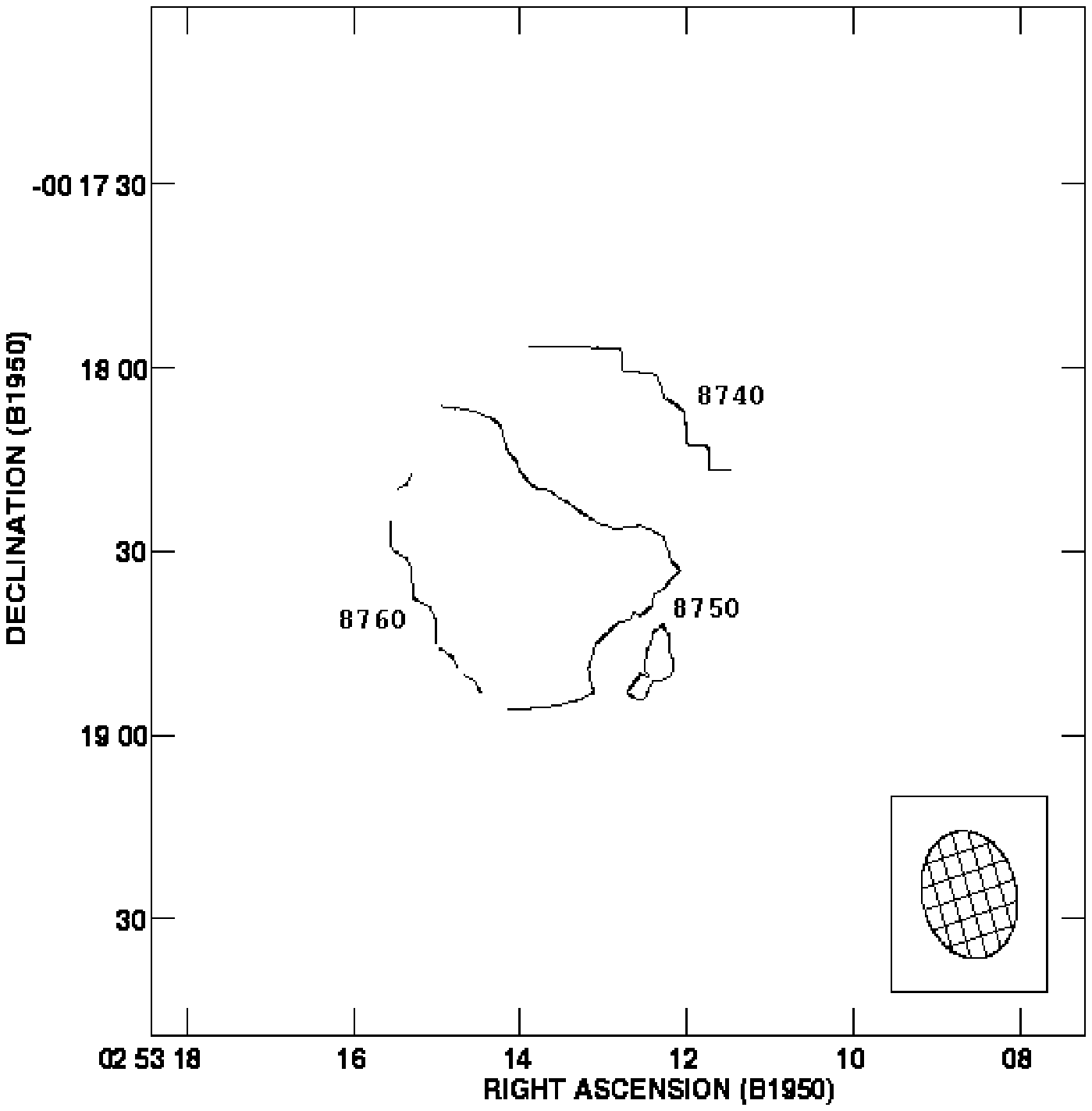]{Mean velocity field of dwarf companion, with the 
velocity contours labeled in units of km s$^{\rm -1}$.}
\end{figure} 

\end{document}